\documentclass[manuscript]{emulateapj}
\usepackage{longtable,amsmath}

\shorttitle{Kinematics of M31 dSphs}
\shortauthors{Collins et al.}
\def\ltsima{$\; \buildrel < \over \sim \;$}
\def\lta{\lower.5ex\hbox{\ltsima}}
\def\gtsima{$\; \buildrel > \over \sim \;$}
\def\simgt{\lower.5ex\hbox{\gtsima}}
\def\kms{{\rm\,km\,s^{-1}}}

\def\kpc{{\rm\,kpc}}

\def\msun{{\rm\,M_\odot}}
\def\lsun{{\rm\,L_\odot}}

\def\pc{{\rm\,pc}}

\newcommand{\Lagr}{\mathcal{L}}

\def\vmaxco{$V_{\rm max}=15.6^{+1.5}_{-1.3}\kms$}
\def\vmaxca{$V_{\rm max}=12.8^{+1.3}_{-1.1}\kms$}

\def\vmaxcmw{$V_{\rm max}=16.2^{+2.8}_{-2.1}\kms$}

\def\vmaxno{$V_{\rm max}=16.2^{+2.6}_{-1.7}\kms$}
\def\vmaxna{$V_{\rm max}=12.8^{+1.4}_{-1.2}\kms$}

\def\vmaxnmw{$V_{\rm max}=18.4^{+2.9}_{-3.1}\kms$}

\def\vmc{$13.5^{+1.1}_{-0.9}$}
\def\vmco{$15.6^{+1.5}_{-1.3}$}
\def\vmca{$12.8^{+1.3}_{-1.1}$}
\def\vmcoa{$15.7^{+2.1}_{-1.7}$}
\def\vmcmw{$16.2^{+2.8}_{-2.1}$}
\def\vmcomw{$15.9^{+3.1}_{-2.1}$}
\def\vmn{$13.6^{+1.3}_{-1.0}$}
\def\vmno{$16.2^{+2.6}_{-1.7}$}
\def\vmna{$12.8^{+1.4}_{-1.2}$}
\def\vmnoa{$16.7^{+3.5}_{-2.4}$}
\def\vmnmw{$18.4^{+2.9}_{-3.1}$}
\def\vmnomw{$18.7^{+4.9}_{-4.1}$}

\def\rsco{$R_S=225^{+70}_{-55}{\rm\,pc}$}
\def\rsca{$R_S=142^{+72}_{-53}{\rm\,pc}$}

\def\rscmw{$R_S=253^{+143}_{-99}{\rm\,pc}$}

\def\rsno{$R_S=664^{+412}_{-232}{\rm\,pc}$}
\def\rsna{$R_S=322^{+247}_{-143}{\rm\,pc}$}

\def\rsnmw{$R_S=1034^{+1508}_{-524}{\rm\,pc}$}

\def\rc{$165^{+58}_{-47}$}
\def\rco{$225^{+70}_{-55}$}
\def\rca{$142^{+72}_{-53}$}
\def\rcoa{$257^{+108}_{-88}$}
\def\rcmw{$253^{+143}_{-99}$}
\def\rcomw{$208^{+119}_{-82}$}
\def\rn{$408^{+221}_{-143}$}
\def\rno{$664^{+412}_{-232}$}
\def\rna{$322^{+247}_{-143}$}
\def\rnoa{$790^{+828}_{-349}$}
\def\rnmw{$1034^{+1508}_{-524}$}
\def\rnomw{$708^{+1816}_{-391}$}

\def\sigco{$\sigma_{V_{\rm max}}=2.8^{+0.5}_{-0.4}\kms$}
\def\sigca{$\sigma_{V_{\rm max}}=3.8^{+0.7}_{-0.6}\kms$}

\def\sigcmw{$\sigma_{V_{\rm max}}=2.9^{+0.8}_{-0.6}\kms$}

\def\signo{$\sigma_{V_{\rm max}}=2.9^{+0.5}_{-0.4}\kms$}
\def\signa{$\sigma_{V_{\rm max}}=3.9^{+0.7}_{-0.6}\kms$}

\def\signmw{$\sigma_{V_{\rm max}}=2.9^{+0.7}_{-0.5}\kms$}

\def\scall{$3.5^{+0.5}_{-0.4}$}
\def\sco{$2.8^{+0.5}_{-0.4}$}
\def\sca{$3.8^{+0.7}_{-0.6}$}
\def\scoa{$3.2^{+0.7}_{-0.6}$}
\def\scmw{$2.9^{+0.8}_{-0.6}$}
\def\scomw{$2.5^{+0.8}_{-0.6}$}
\def\sn{$3.6\pm0.5$}
\def\sno{$2.9^{+0.5}_{-0.4}$}
\def\sna{$3.9^{+0.7}_{-0.6}$}
\def\snoa{$3.2^{+0.7}_{-0.6}$}
\def\snmw{$2.9^{+0.7}_{-0.5}$}
\def\snomw{$2.4^{+0.7}_{-0.5}$}

\def\s{\ifmmode \widetilde \else \~\fi}
\def\={\overline}

\def\spose#1{\hbox to 0pt{#1\hss}}

\def\lta{\mathrel{\spose{\lower 3pt\hbox{$\mathchar"218$}}
     \raise 2.0pt\hbox{$\mathchar"13C$}}}
\def\gta{\mathrel{\spose{\lower 3pt\hbox{$\mathchar"218$}}
     \raise 2.0pt\hbox{$\mathchar"13E$}}}
\def\Dt{\spose{\raise 1.5ex\hbox{\hskip3pt$\mathchar"201$}}}    
\def\dt{\spose{\raise 1.0ex\hbox{\hskip2pt$\mathchar"201$}}}    

\shorttitle{The Masses of Local Group dSphs}
\shortauthors{Collins et al.}

\begin{document}
\title{The masses of Local Group dwarf spheroidal galaxies: The death of the
  universal mass profile}

\author{Michelle L. M. Collins\altaffilmark{1}, Scott C. Chapman\altaffilmark{2,3}, R. M. Rich\altaffilmark{4}, Rodrigo A. Ibata\altaffilmark{5}, Nicolas F. Martin\altaffilmark{5,1}, Michael J. Irwin\altaffilmark{2}, Nicholas F. Bate\altaffilmark{6}, Geraint F. Lewis\altaffilmark{6}, Jorge Pe\~narrubia\altaffilmark{7}, Nobuo Arimoto\altaffilmark{8,9}, Caitlin M. Casey\altaffilmark{10}, Annette M. N. Ferguson\altaffilmark{11}, Andreas Koch\altaffilmark{12}, Alan W. McConnachie\altaffilmark{13}, Nial Tanvir\altaffilmark{14}}
\altaffiltext{1}{Max-Planck-Institut
  f\"ur Astronomie, K\"onigstuhl 17, D-69117 Heidelberg, Germany}
\altaffiltext{2}{Institute of
  Astronomy, Madingley Rise, Cambridge, CB3 0HA ,UK}
\altaffiltext{3}{Dalhousie University Dept. of Physics and Atmospheric Science
  Coburg Road Halifax, B3H1A6, Canada}
\altaffiltext{4}{Department of Physics and Astronomy, University of
  California, Los Angeles, CA 90095-1547}
\altaffiltext{5}{Observatoire de Strasbourg,11, rue de l'Universit\'e,
  F-67000, Strasbourg, France}
\altaffiltext{6}{Sydney Institute
  for Astronomy, School of Physics, A28, University of Sydney, NSW 2006,
  Australia}
\altaffiltext{7}{Ram\'on y Cajal Fellow, Instituto de
  Astrof\'isica de Andalucia-CSIC, Glorieta de la Astronom\'ia s/n, 18008,
  Granada, Spain}
\altaffiltext{8}{Subaru Telescope, National Astronomical Observatory of Japan 650 North
  A'ohoku Place, Hilo, Hawaii 96720, U.S.A.}
\altaffiltext{9}{Graduate University for Advanced Studies 2-21-1 Osawa, Mitaka,
  Tokyo 181-8588, Japan}
\altaffiltext{10}{Institute for Astronomy, 2680 Woodlawn Drive Honolulu, HI 96822-1839
  USA }
\altaffiltext{11}{Institute for Astronomy, University of Edinburgh, Royal
  Observatory, Blackford Hill, Edinburgh, EH9 3HJ, UK}
\altaffiltext{12}{Zentrum f\"ur Astronomie der Universit\"at Heidelberg, Landessternwarte,
  K\"onigstuhl 12, 69117 Heidelberg, Germany}
\altaffiltext{13}{NRC Herzberg Institute of Astrophysics, 5071 West Saanich
  Road, British Columbia, Victoria V9E 2E7, Canada}
\altaffiltext{14}{Department of
  Physics \& Astronomy, University of Leicester, University Road, Leicester
  LE1 7RH, UK}

\begin{abstract}
  We investigate the claim that all dwarf spheroidal galaxies (dSphs) reside
  within halos that share a common, universal mass profile as has been derived
  for dSphs of the Galaxy. By folding in kinematic information for 25
  Andromeda dSphs, more than doubling the previous sample size, we find that a
  singular mass profile can not be found to fit all the observations
  well. Further, the best-fit dark matter density profile measured for solely
  the Milky Way dSphs is marginally discrepant (at just beyond the 1$\sigma$
  level) with that of the Andromeda dSphs, where a profile with lower maximum
  circular velocity, and hence mass, is preferred. The agreement is
  significantly better when three extreme Andromeda outliers, And XIX, XXI and
  XXV, all of which have large half-light radii ($\gta600$pc) and low velocity
  dispersions ($\sigma_v<5\kms$) are omitted from the sample. We argue that
  the unusual properties of these outliers are likely caused by tidal
  interactions with the host galaxy. 
\end{abstract}


\section{Introduction}

Within the accepted cosmological paradigm -- $\Lambda$ cold dark matter ($\Lambda$CDM) --
approximately 85\% of the matter in the Universe is thought to be dark
\citep{komatsu11,planck1,planckcp}. As such, understanding the nature of this component
is of the upmost importance. While the precise properties of this exotic
matter are still unknown, various predictions about its behaviour and mass
distribution within galaxies have been made by both cosmological and particle
physics models.

While it has experienced great success on large scales, this cosmological
model has run into some difficulty adequately explaining a number of
observations made on smaller scales. In particular, a number of mismatches
between observation and theory with regard to the smallest, most dark matter
dominated galaxies we are able to observe - the dwarf spheroidals (dSphs) -
have been defined. The missing satellite problem has floated around for some
time now \citep{klypin99,moore99} and refers to the dearth of observed
luminous subhalos around the Milky Way (MW) and Andromeda (M31), compared to
the vast number of dark matter subhalos seen within dark matter only
simulations. The scope of this problem has somewhat lessened over the years,
as the community seems largely satisfied that this can be resolved with future
observations and a better understanding of the physics underlying galaxy
formation. Firstly, we do not expect all subhalos seen within the simulations
to have enough mass to accrete and retain the gas required to efficiently form
stars. As such, a lower mass of $V_{\rm max}\sim10-15\kms$ is placed on
luminous galaxy formation \citep{penarrubia08b,koposov09}. Secondly, our
observations are currently incomplete (both areally and in terms of surface
brightness). By considering and correcting for the completeness of current
surveys of the halos of these galaxies
(\citealt{tollerud08,koposov09,walsh09}; Martin et al, in prep.) the number of
observed vs. predicted subhalos can be brought into much better agreement.

Another observed and as yet unresolved tension is the ongoing `cusp-core'
debate, which refers to the shape of the dark matter profile of galactic halos
as radius tends to zero. With central mass to light ratios of typically
$M/L>10\msun/\lsun$ (e.g \citealt{mateo98,walker09b,tollerud12,collins13}),
one can treat the stars contained within dSphs as massless tracers of the dark
matter potential, and their small scales (half-light radii of
$r_{\rm{half}}\sim100-1000$s pc) allow us to probe their mass profiles in the
very centers of their halos. This allows us to test the predictions from
cosmological dark matter only simulations of halo mass profiles; namely that
these are steeply cusped (i.e., the density dramatically increases for
decreasing radius, \citealt{navarro97}). Increasingly, observations of dwarf
spheroidals (and other low surface brightness galaxies) show evidence for
constant density cores in the centers of galaxy halos
(e.g. \citealt{deblok02,deblok03,deblok05,walker11,amorisco12,jardel12}). Whether
this tension can be resolved by appealing to baryonic processes, such as
feedback from star formation or tidal stripping is something that is currently
being debated (e.g. \citealt{zolotov12,brooks12,garrison13}) and a theme we
shall return to later on.

Related to the cusp-core problem is the `too big to fail' (TBTF) problem,
which was originally identified by \citet{read06b} and has received much
attention recently from others (e.g. \citealt{kolchin11,kolchin12}). With the
limited kinematic data currently available for dSph galaxies, it is not
possible to accurately measure the slopes of their density profiles in many
cases, but from measurements of their central velocity dispersion, $\sigma_v$,
one can get a good grasp on the central masses, i.e. the mass within the
2-dimensional half-light radius, $r_{\rm half}$, of these systems
\citep{walker09b,wolf10}, and compare these with those of simulated
subhalos. Such an exercise was undertaken by \citet{kolchin11} using the
Aquarius set of simulations \citep{springel08}, and they found that each
MW-like Aquarius halo they studied had of order 10 subhalos with central
masses that were significantly higher than those of the MW dSphs. This means
either the most massive subhalos within MW systems do not necessarily form
stars, or that we are missing some crucial physics from these models (either
baryonic, or with respect to the properties of dark matter itself) that can
explain this discrepancy.

Each of these problems are currently being investigated by observers and
theorists alike, with proposed solutions ranging from fiddling with baryonic
physics (star formation, feedback, tidal forces etc.) to redefining the
cosmological paradigm (e.g. modified Newtonian dynamics, warm dark matter,
self interacting dark matter). From the observers point of view, one obvious
avenue has been to extend our sample of objects whose kinematics are well
measured. Obtaining the necessary kinematic information with which to study
the central masses of dSphs (i.e. radial velocities of individual stars within
these systems) is exceptionally challenging, meaning the majority of studies
thus far have focused on the 20 surrounding our own Galaxy. It has only been
within the last decade that telescopes capable of measuring kinematics of
extragalactic dSphs have become available. Now, thanks to several recent
papers
(e.g. \citealt{kalirai10,collins10,collins11b,collins13,tollerud12,tollerud13}),
we can add to this sample a further 25 dSphs from the Andromeda system, more
than doubling our sample size. Two of these works in particular, Tollerud et
al. (2012, henceforth T12) and Collins et al. (2013, henceforth C13) have
demonstrated that the majority of the M31 systems have very similar central
masses to their MW counterparts, which would imply that the self-same tensions
discussed above for the MW also apply to the M31 dSph system. In addition,
they highlighted a number of M31 dSphs whose masses appear lower than would be
expected when comparing with expectations based on Milky Way dSphs, casting
some doubt on the notion that all dSph galaxies are hosted within dark matter
halos whose central mass profiles are universal, i.e., behave in a
statistically similar way as a function of radius
\citep{mateo98,strigari08,walker09b,wolf10}.

In this work, we revisit the idea of universal mass profile of
\citet{walker09b} for the dSph population by including the M31 objects into
the analysis. In \S~\ref{sect:results} we show that a singular mass profile,
be it a \citet{navarro97} cusp (NFW) or a constant density core, provides a
poor fit to the Local Group dSphs, and instead we advocate a statistical range
of best fit mass profiles that track the scatter in mass for a given
half-light radius in this population. We then compare these findings with
numerical simulations, demonstrating that the mismatches discussed above do
not simply go away with a larger sample of systems. We identify a number of
unusual systems in M31 whose masses may pose a challenge to our understanding
of galaxy formation and evolution in \S~\ref{sect:outliers}. Finally, in
\S~\ref{sect:discussion}, we go on to discuss how the proposed solutions to
these problems stack up to the observations, before we conclude in
\S~\ref{sect:conclusions}.

\begin{figure}
  \begin{center}
     \includegraphics[angle=0,width=0.9\hsize]{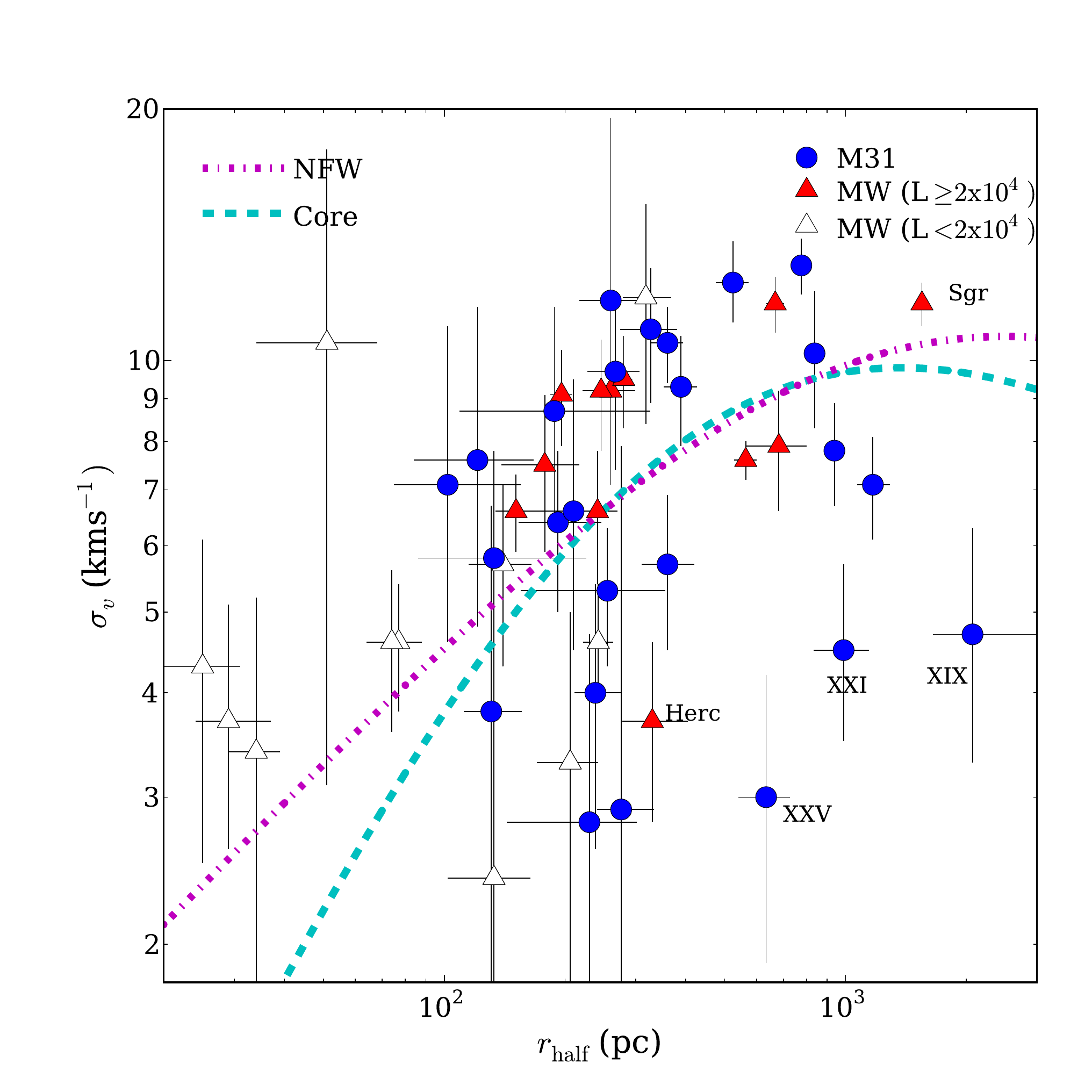}
     \caption{Half-light radius vs. velocity dispersion for MW (red triangles)
       and M31 dSphs (blue circles). Overlaid are best fitting NFW and core
       mass profiles to these data. Open symbols represent MW dSphs that are
       too faint to be observed in M31, and hence are excluded from the
       fits. $\sim50\%$ of all observations are inconsistent with these fits,
       undermining the notion that all dSphs are embedded in halos that follow
       a universal density profile.}
  \label{fig:universal}
  \end{center}
\end{figure}

\section{Data}
\label{sect:data}

As dSphs are largely dispersion supported systems with little or no evidence
of rotation, their velocity dispersions, in combination with their half-light
radii, can be utilized to estimate their central masses. For the Milky Way
population, we rely on the compilation of kinematic and structural properties
formed by \citet{walker09b}, although we exclude the tidally disrupting
Sagittarius galaxy (Sgr) from further analysis, as it is currently not in
equilibrium.  The compilation from \citet{walker09b} were used within that
work to define the universal mass profile for dSph galaxies, which we shall
discuss further below. Since then, three Galactic dSphs have benefitted from
further study of their kinematics; Hercules \citep{aden09}, B\"ootes I
\citep{koposov11} and Segue 2 \citep{kirby13}. In all cases, the velocity
dispersions (and hence, calculated masses) have reduced.

For the Andromeda dSphs, we take our kinematics and structural properties from
the final table in C13, which is a compilation of the best velocity
dispersions from that work (those of Andromeda VI, XI, XVII, XIX, XX, XXII,
XXIII, XXV, XXVI, XXVIII and XXX [Casseopia II]), and from those presented by
T12 (Andromeda I, III, V, VII, IX, X, XIII, XIV, XV and XVIII). For And XVI
and XXI, we use newly derived values for the velocity dispersions of these
objects that have been made from much larger samples of member stars
($\sigma_v=5.6\pm1.0$ and $\sigma_v=5.4\pm0.9$ for XVI and XXI respectively,
Collins et al. in prep). The velocity dispersion for Andromeda (And) II is
taken from \citet{ho12}, and that of And XXIX is taken from
\citet{tollerud13}. The velocity dispersion for And XII is unresolved, so we
omit that from our study here. Owing to difficult observing conditions, the
velocity dispersion of And XXIV is not well constrained, so we omit this value
too. Finally, as And XXVII is likely a heavily disrupted system whose
kinematics and structural properties are not well constrained (C13, Martin et
al., in prep), we also remove this system from our analysis. This leaves us
with a sample of 25 M31 dSph galaxies for which velocity dispersions have been
reliably estimated. The structural properties are taken from
\citet{mcconnachie12} and Martin et al. (in prep., for the M31 dwarf galaxies that fall
in the PAndAS footprint), updated based on the
revised distances to the Andromeda dSphs presented in \citet{conn12b}.

\section{Results}
\label{sect:results}

\subsection{A universal mass profile?}
\label{sect:ump}

\begin{figure*}
  \begin{center}
     \includegraphics[angle=0,width=0.3\hsize]{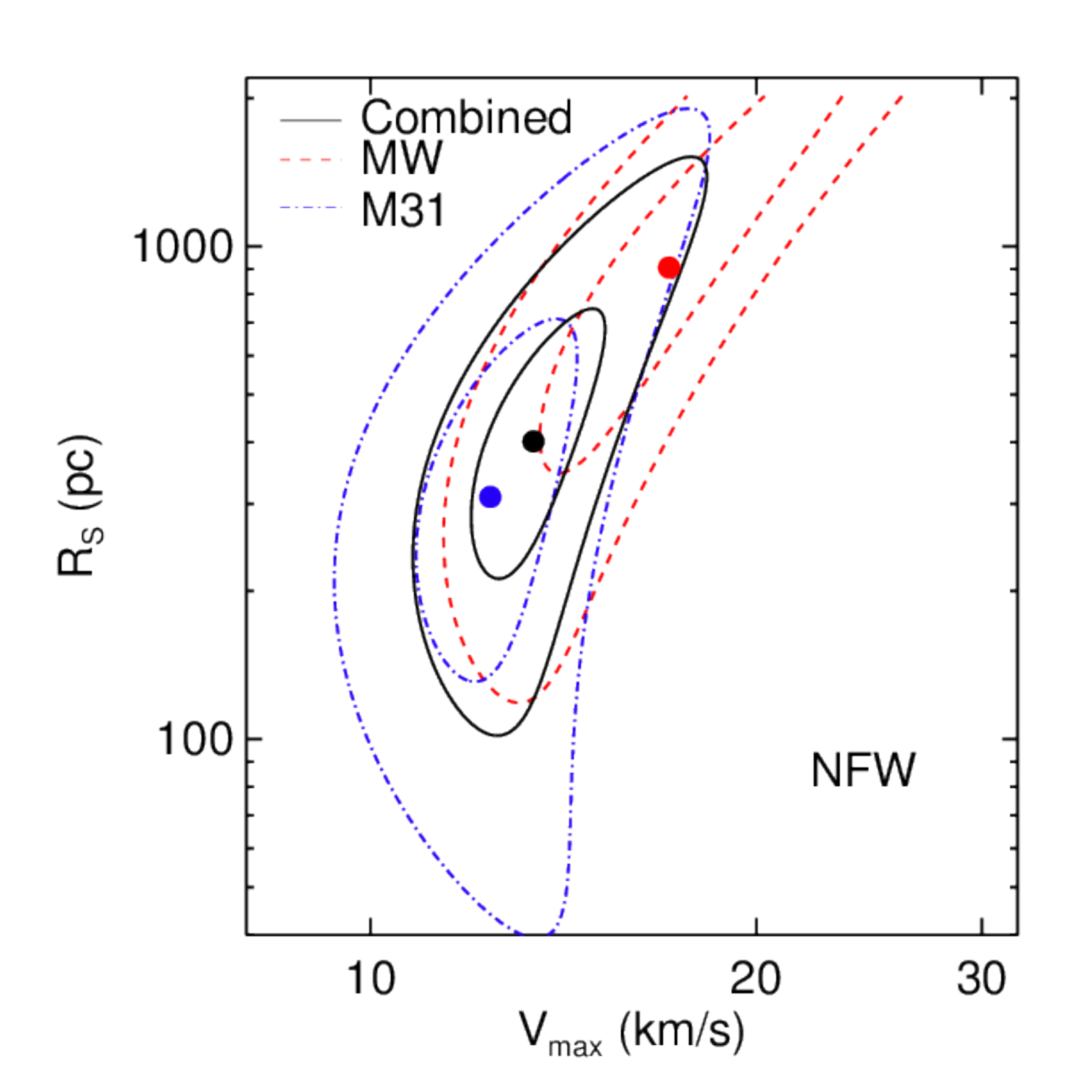}
    \includegraphics[angle=0,width=0.3\hsize]{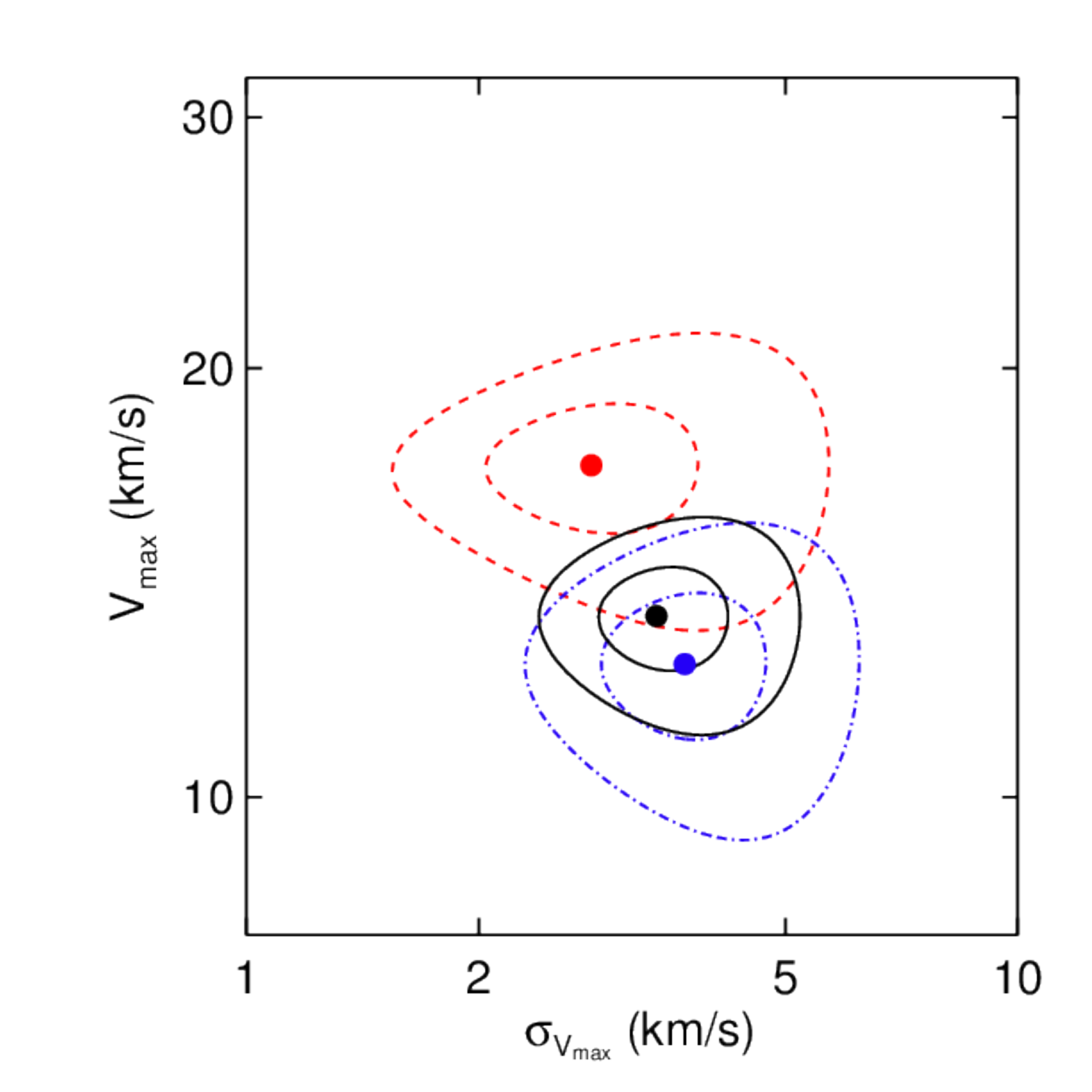}
     \includegraphics[angle=0,width=0.3\hsize]{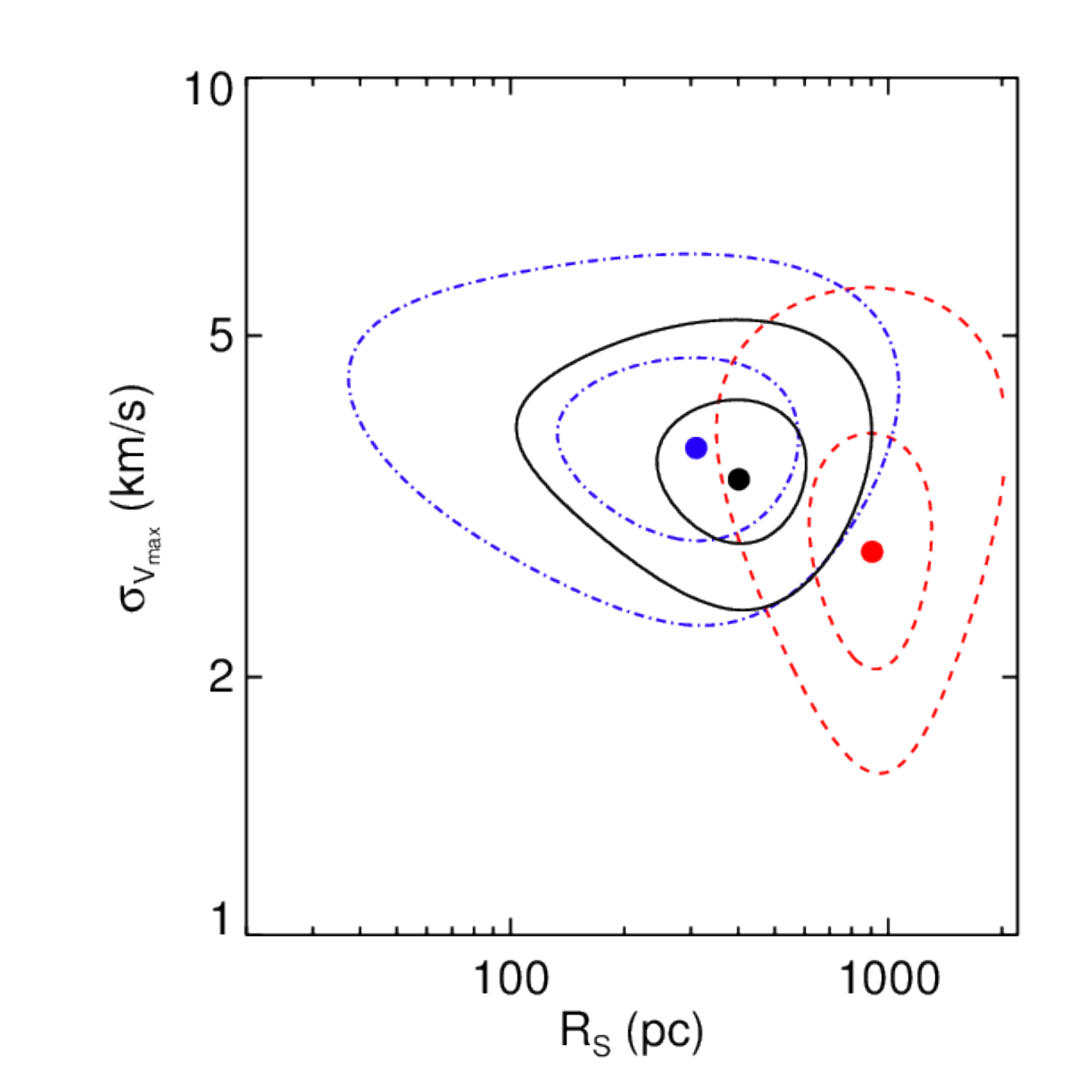}
     \includegraphics[angle=0,width=0.3\hsize]{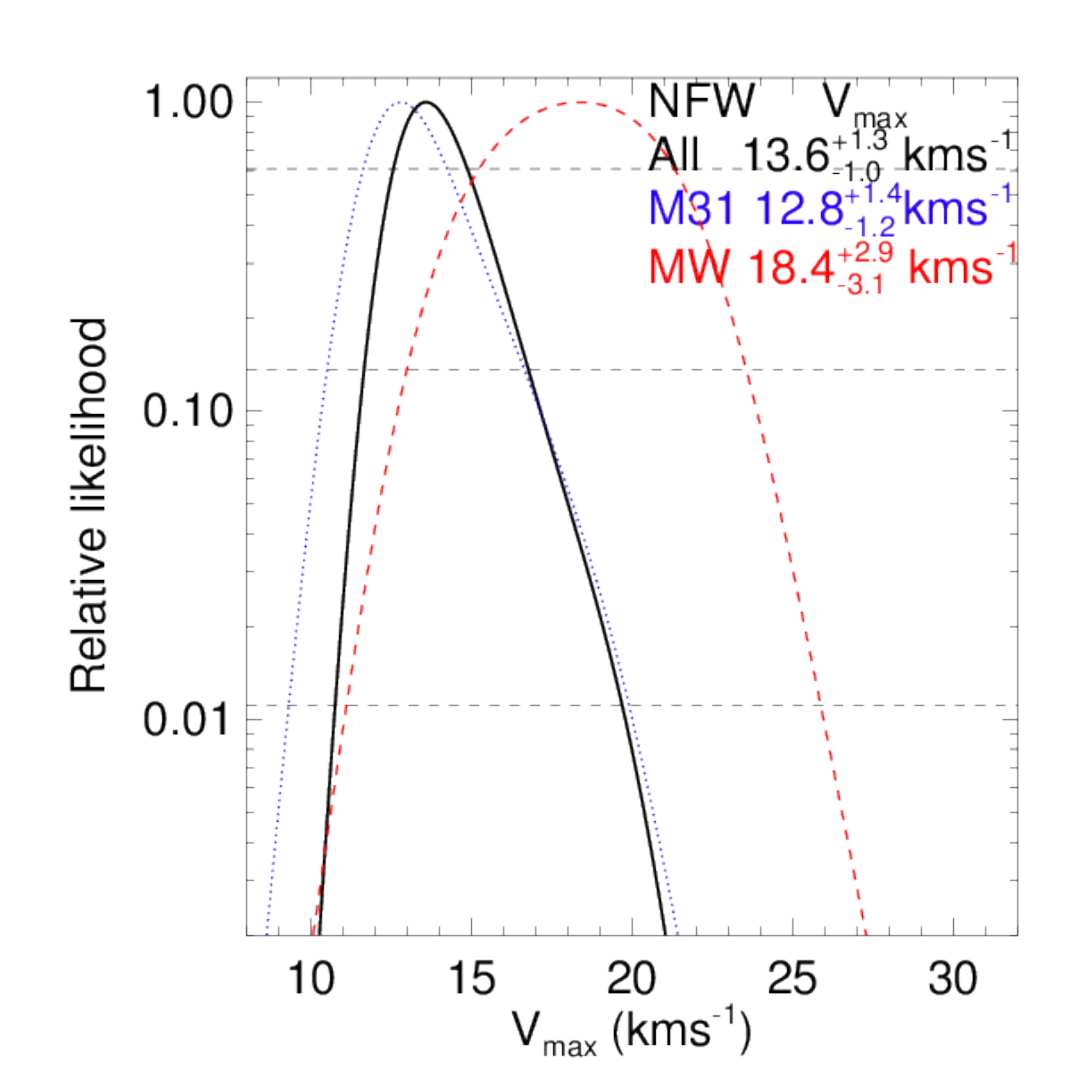}
    \includegraphics[angle=0,width=0.3\hsize]{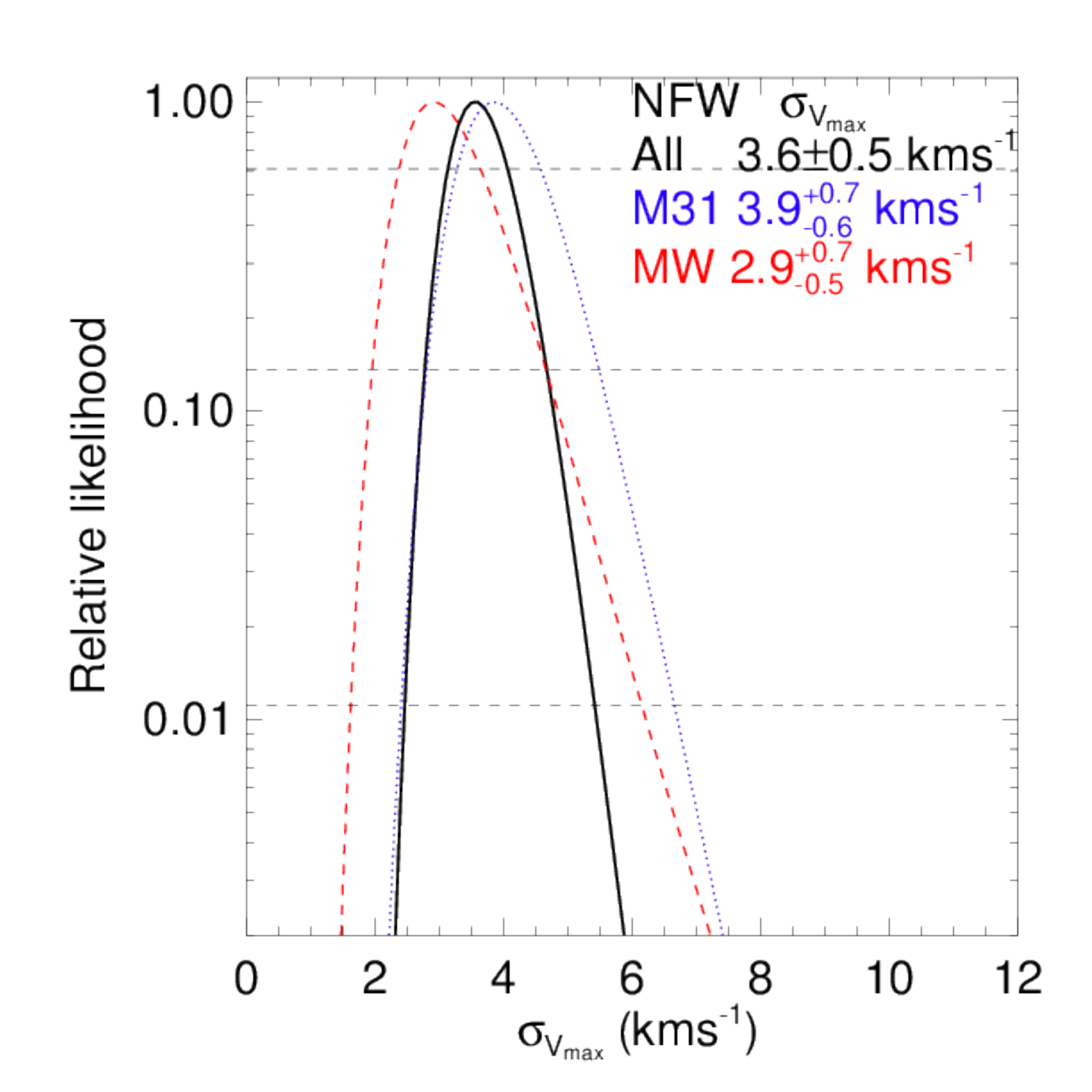}
     \includegraphics[angle=0,width=0.3\hsize]{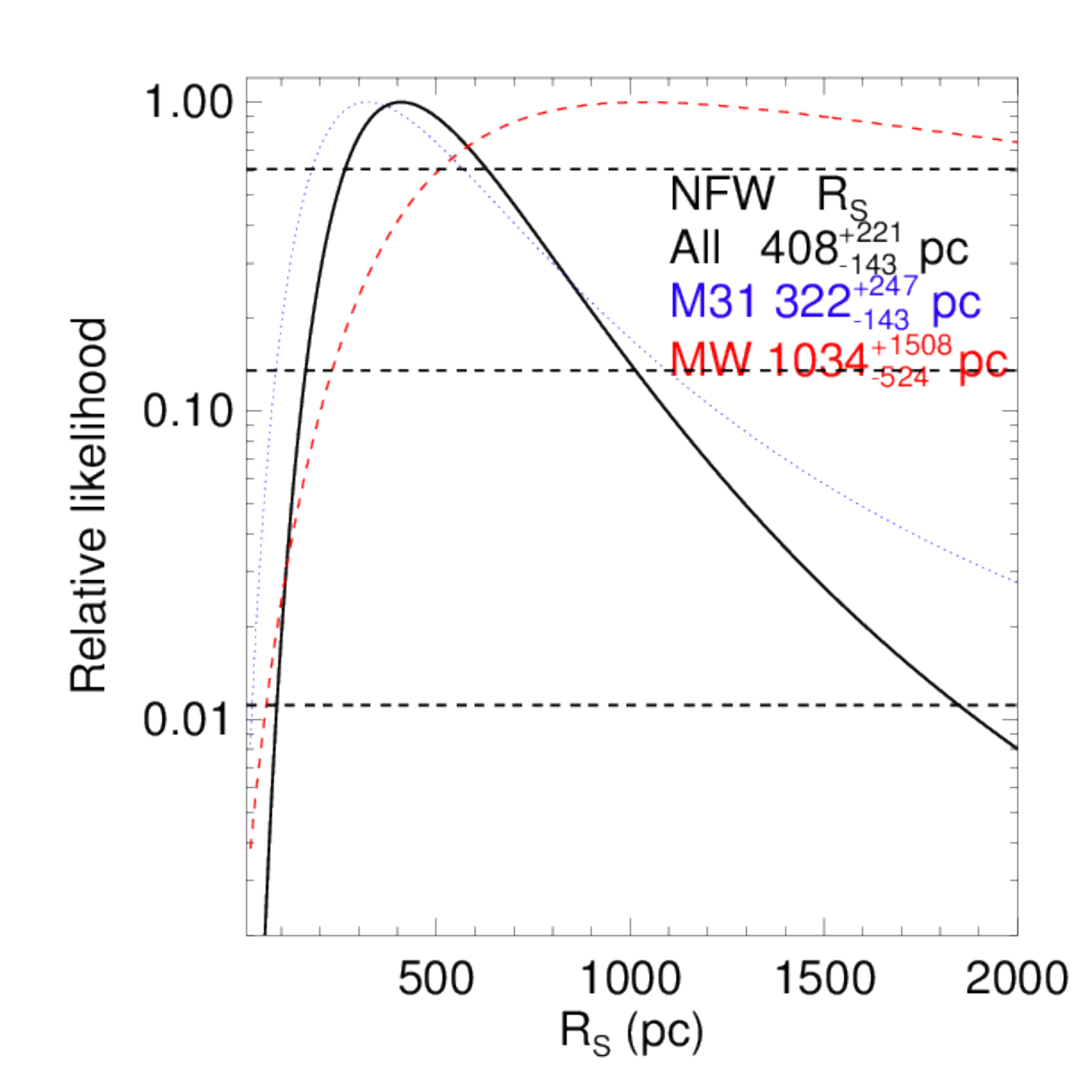}
      \includegraphics[angle=0,width=0.3\hsize]{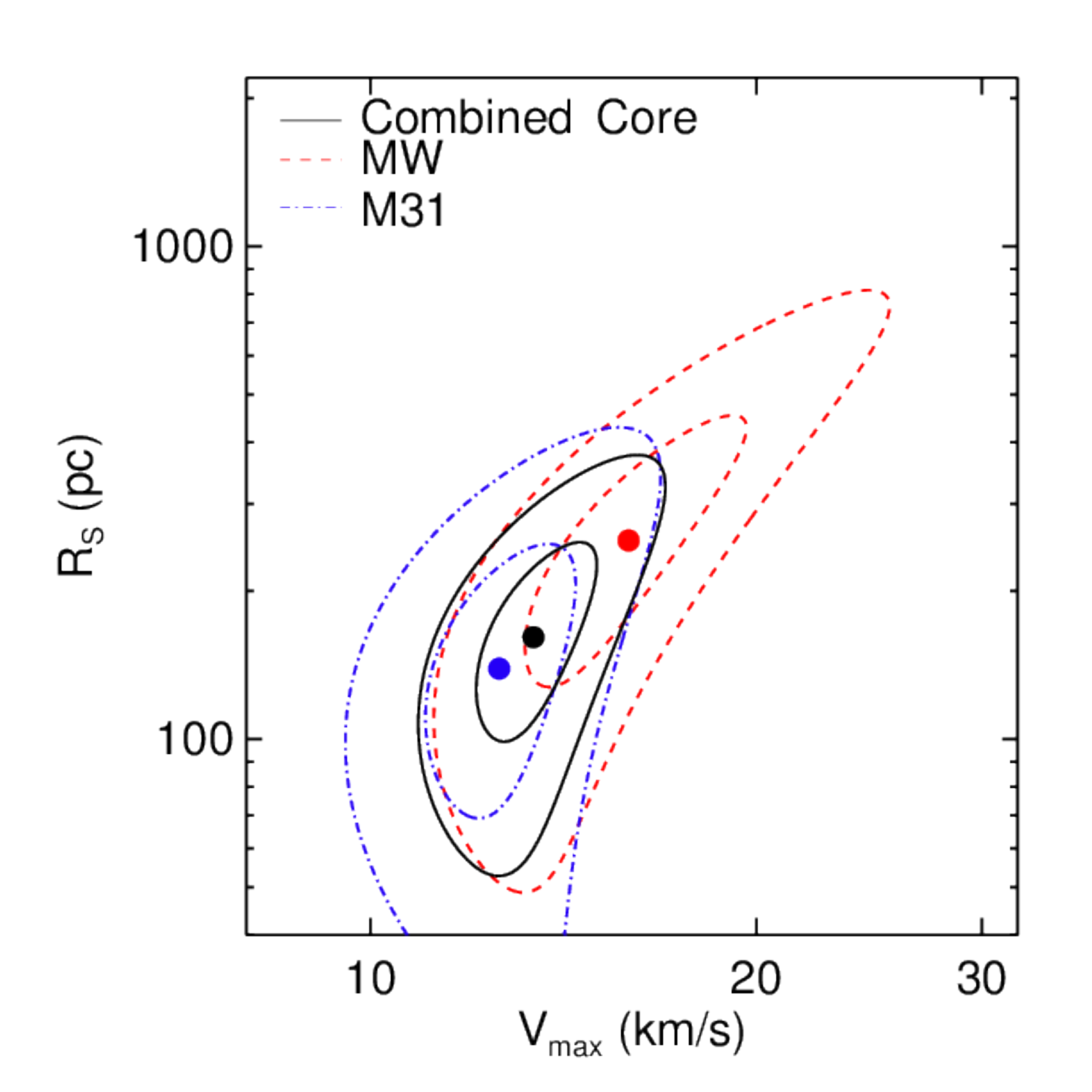}
    \includegraphics[angle=0,width=0.3\hsize]{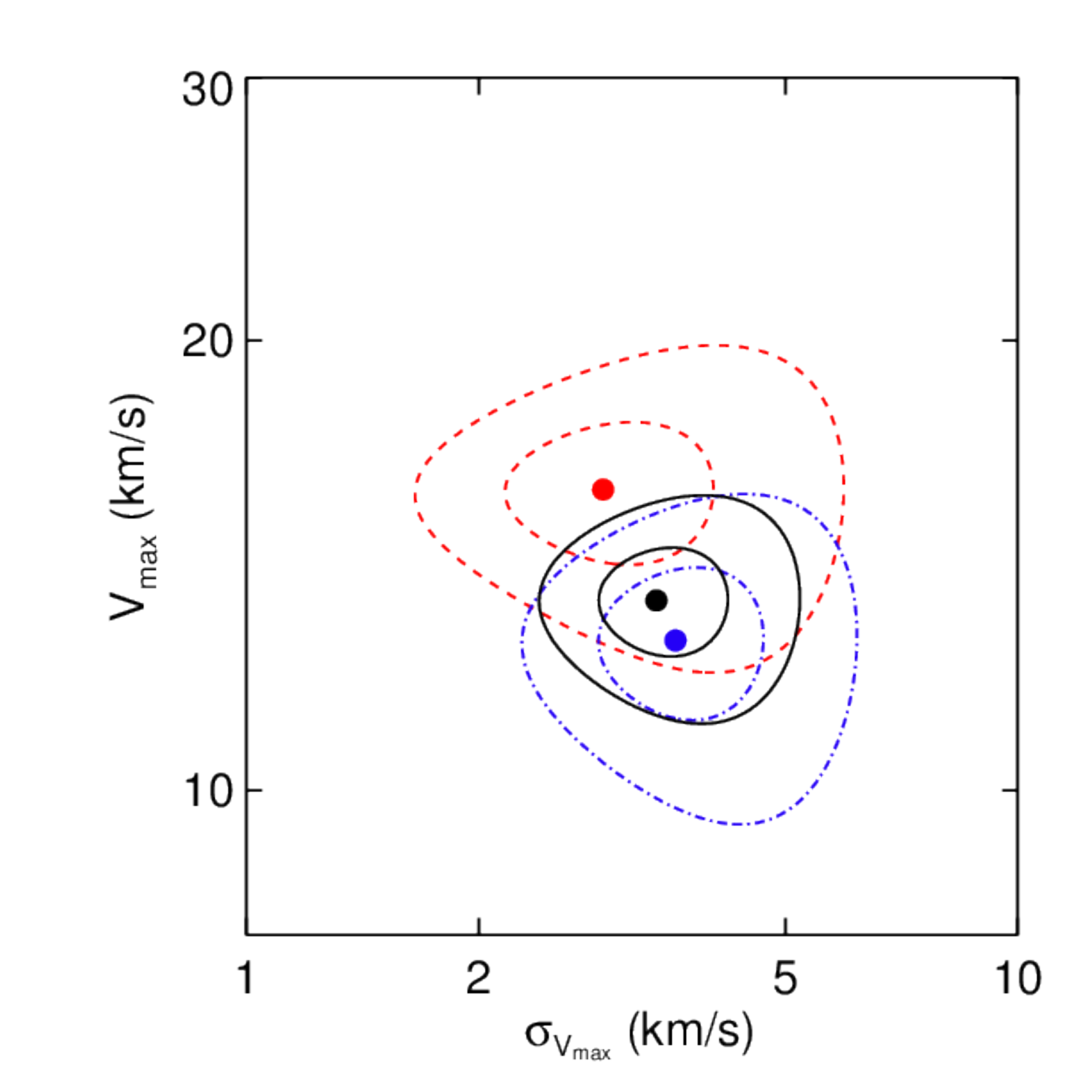}
     \includegraphics[angle=0,width=0.3\hsize]{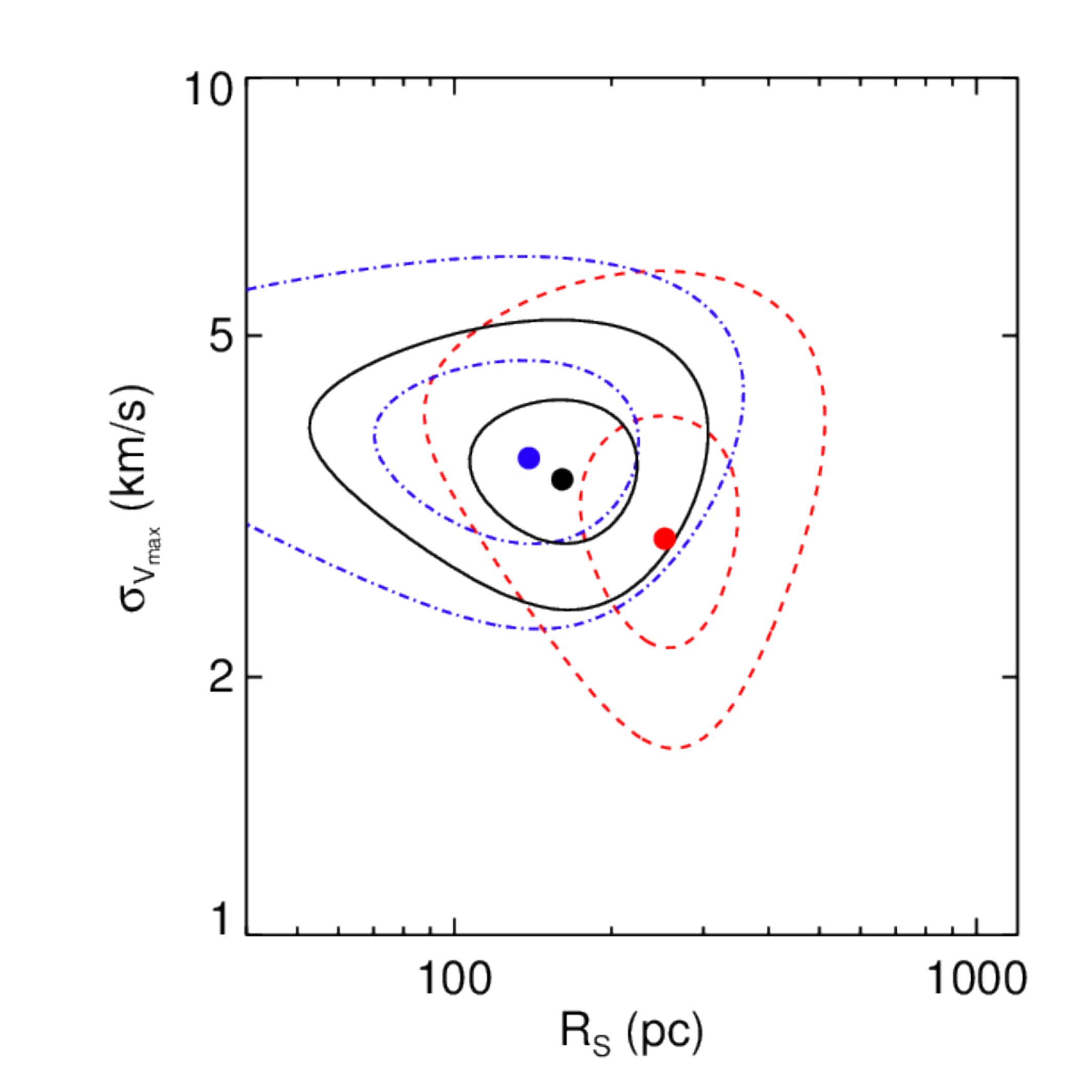}
     \includegraphics[angle=0,width=0.3\hsize]{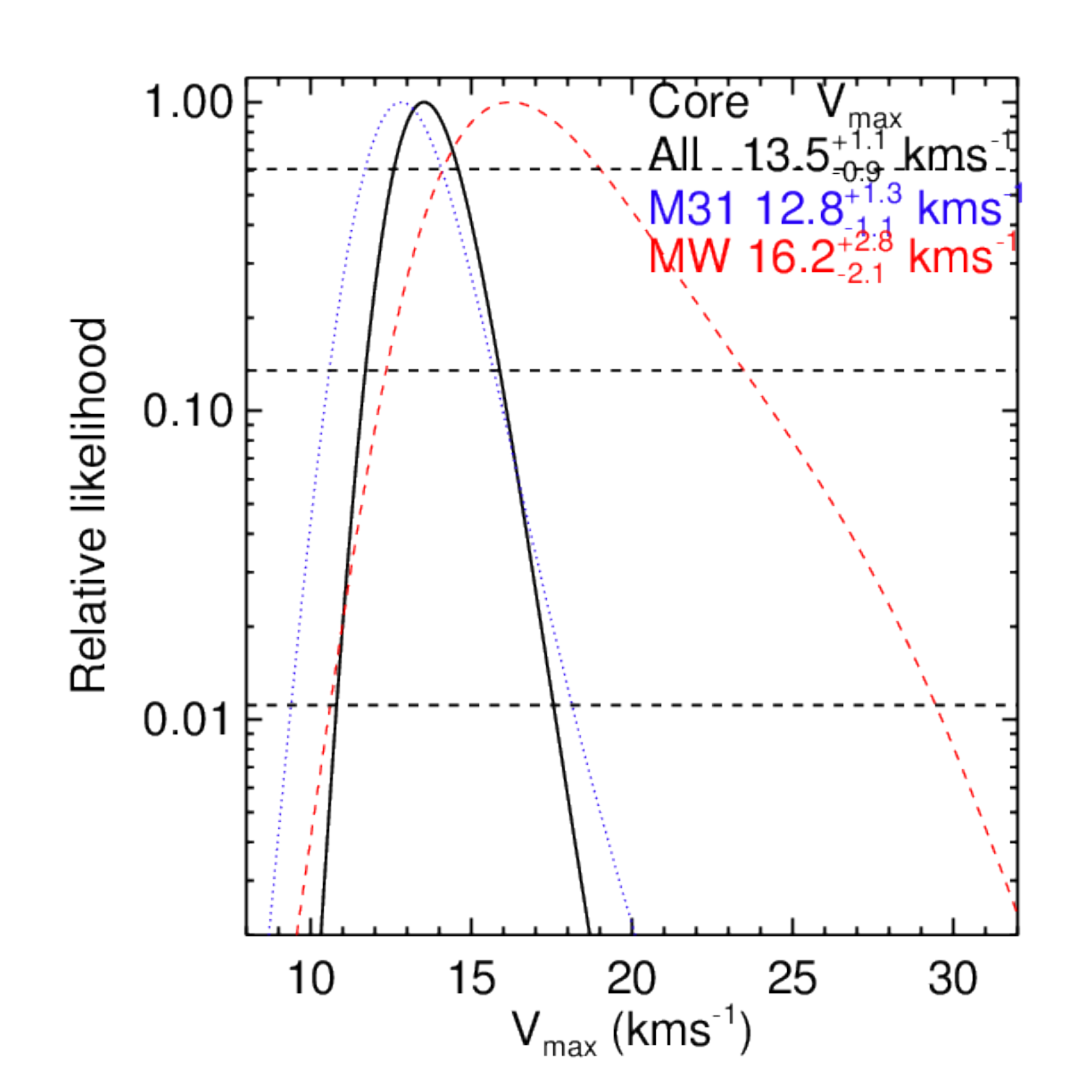}
    \includegraphics[angle=0,width=0.3\hsize]{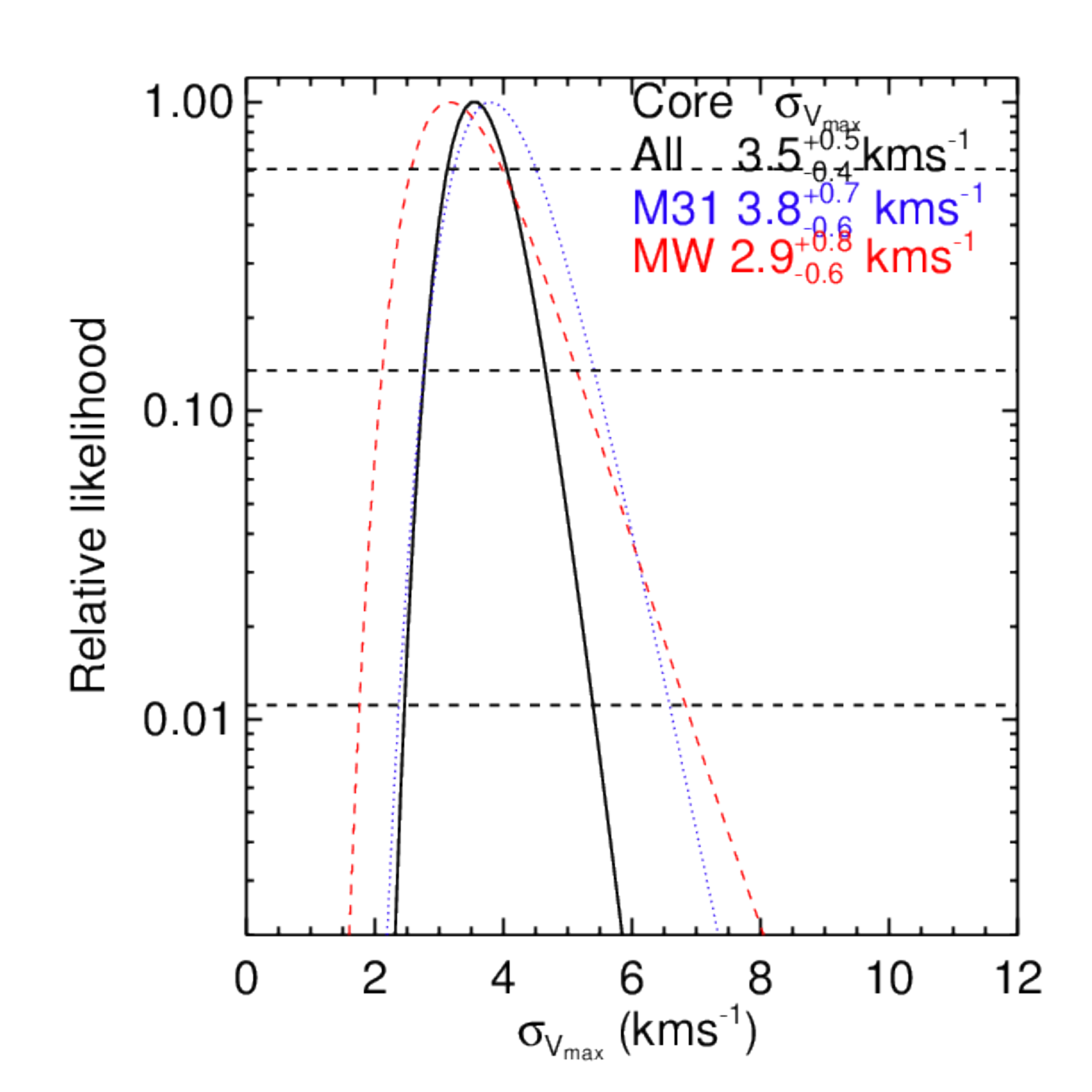}
     \includegraphics[angle=0,width=0.3\hsize]{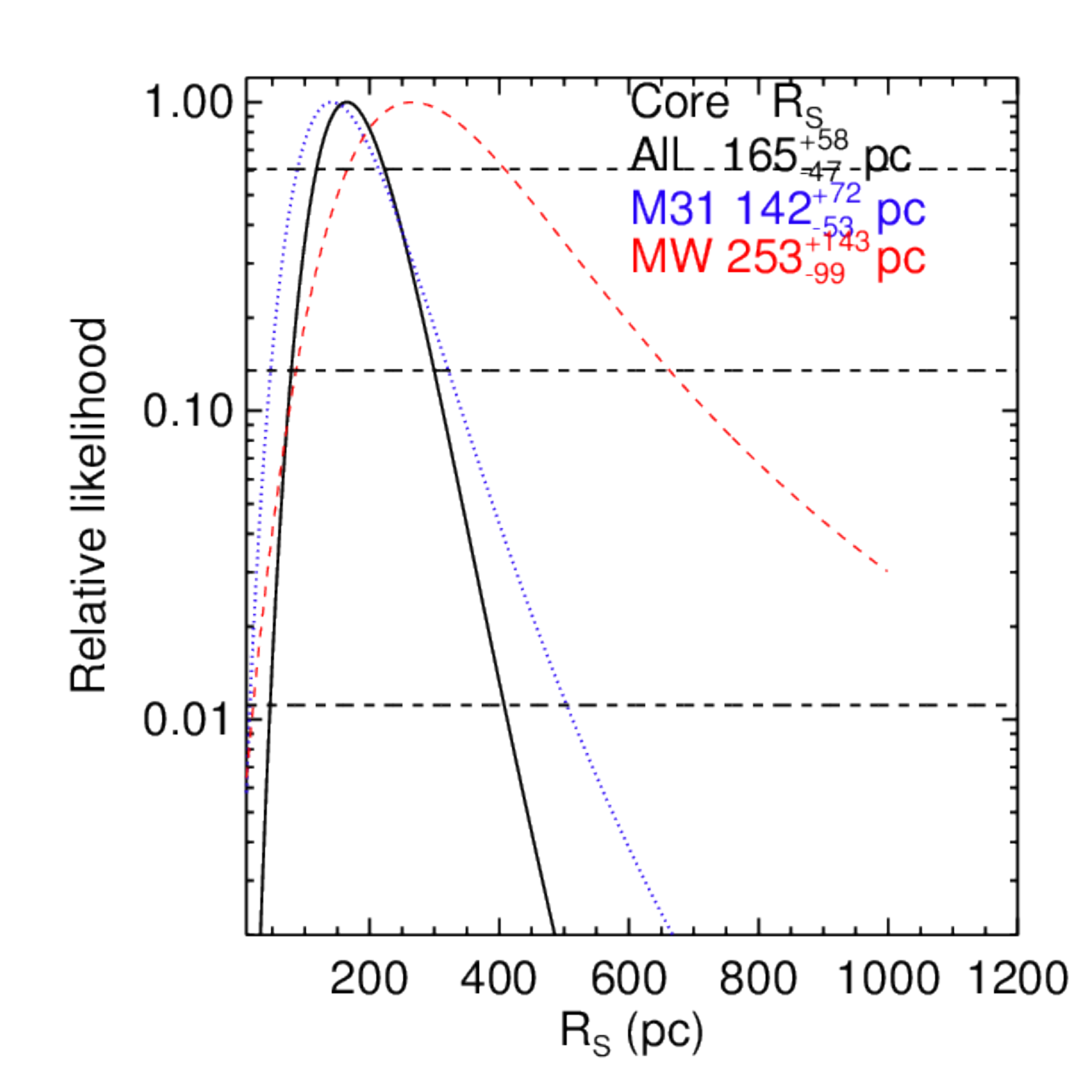}
     \caption{{ \bf Top row:} 2 dimensional likelihood contours for the three
       free parameters ($V_{\rm max}$, $\sigma_{V_{\rm max}}$ and $R_S$) in
       the NFW mass profile fits to the MW dSphs (red dashed contours), M31
       dSphs (blue dot-dashed contours) and all Local Group dSphs (solid black
       contours). The contours represent the 1 and $2\sigma$ (i.e. 68\% and
       95\%) confidence intervals for these values. {\bf Second row:} The
       resulting 1 dimensional marginalized likelihoods for $V_{\rm max}$,
       $\sigma_{V_{\rm max}}$ and $R_S$ (from left to right) for the MW, M31
       and full sample. Horizontal dashed lines represent the 1, 2 and 3$\sigma$
       (i.e. 68\%,95\% and 99.7\%) confidence intervals, derived assuming a
       Gaussian probability distribution. The best fit $\sigma_{V_{\rm max}}$
       and $R_S$ agree quite well between the MW and M31 case, however the
       values of $V_{\rm max}$ for the MW and M31 are marginally inconsistent,
       at the level of $1\sigma$, with the M31 dSphs favoring fits with lower
       central masses. {\bf Third and fourth rows:} As above six panels, but
       for cored density profile fits. Again, the best fit values of $V_{\rm
         max}$ for the MW and M31 are discrepant at the level of $1\sigma$.}
  \label{fig:allML}
  \end{center}
\end{figure*}

We focus our analysis here on the inclusion of the masses of the M31 dSphs
into the universal density profile of \citet{walker09b}. In their work, the
authors were spurred on by the earlier results of \citet{strigari08} that
showed that all of the MW dSphs for which kinematic data were available were
consistent with having the same mass contained within a radius of 300 pc
(roughly $1\times10^7\msun$) despite spanning 6 decades in
luminosity. \citet{strigari08} used this result to argue that it was possible
that all dSphs inhabited a universal dark matter halo, where
the density as a function of radius was identical, irrespective of the number
of stars the halo hosted. However, \citet{wolf10} demonstrated
that extrapolations to both larger and smaller radii than the true half light
radius are extremely uncertain in cases where the velocity anisotropy is
unknown, and this is true for all the Local Group dSphs. For objects with
$r_{\rm half}<<300$~pc one has to extrapolate to radii inhabited by no
tracers, where tidal stripping may have removed the outer dark matter envelope
\citep{penarrubia08b}. That means that for some galaxies extrapolating out to
300~pc could over-estimate the enclosed mass by several orders of
magnitude. In the interest of trying to measure a more meaningful
mass for these objects to determine whether dSphs truly resided within a
universal halo, \citet{walker09b} measured the velocity
dispersion, and hence mass, within the half-light radius of the MW
dSphs. Then, by treating each velocity dispersion measurement from MW dSphs as
a measurement of the velocity dispersion at a given radius (the half-light
radius of the dSph in question) within a single dark matter halo, they could
map out the velocity dispersion profile for this singular halo. In particular,
they tested the cosmologically motivated Navarro Frenk White (NFW,
\citealt{navarro97}) density profile:

\begin{equation}
r_{\rm half}\sigma_v^2=\frac{2\eta
  R_SV_{\mathrm{max}}^2}{5}\times\left[\frac{\mathrm{ln}(1+r_{\mathrm{half}}/R_S)-\frac{r_{\mathrm{half}}/R_S}{1+r_{\mathrm{half}}/R_S}}{\rm{ln}(1+\eta)-\frac{\eta}{1+\eta}}\right],
\label{eqn:nfw}
\end{equation}

\noindent where $V_{\rm max}$ is the maximum circular velocity of the halo,
$R_S$ is the scale radius of the halo and $\eta=2.16$. They also used a cored
density profile where:

\begin{equation}
\begin{aligned}
r_{\mathrm{half}}\sigma_v^2=\frac{2\eta R_SV_{\mathrm{max}}^2}{5(\mathrm{ln}[1+\eta])+\frac{2}{1+\eta}-\frac{1}{2(1+\eta)^2}-\frac{3}{2}}\times\\
\left[\mathrm{ln}(1+r_{\rm half}/R_S)+\frac{2}{1+r_{\rm half}/R_S}-\frac{1}{2(1+r_{\rm half}/R_S)^2}-\frac{3}{2}\right],
\end{aligned}
\label{eqn:core}
\end{equation}

\noindent with $\eta=4.42, \alpha=1$ and $\gamma=0$. The results of this study
showed that the MW dSphs were consistent with having formed with a universal
mass profile, although the authors noted that there was significant
scatter about this relation, a factor of 2 greater than expected from the
observational uncertainties alone. Later that same year, a revised study of the mass
of the Hercules dSph \citep{aden09}, which provided a better treatment of the
contaminating foreground population, determined a much lower value for the
velocity dispersion of this object ($3.72\pm0.91\kms$ vs. $5.1\pm0.9\kms$
from \citealt{simon07}). With their revised value, they showed that the mass
of Hercules was not consistent with the universal mass profile. As Hercules is
likely significantly affected by tides \citep{aden09,martin10}, this is
perhaps not unexpected. Similarly, an analysis of the B\"ootes I dSph by
\citet{koposov11}, who used multi-epoch observations taken with the VLT and
implemented an enhanced data reduction approach to measure extremely precise
radial velocities, measured a velocity dispersion of
$\sigma_v=4.6^{+0.8}_{-0.6}\kms$, significantly lower than that of
$\sim6.5\kms$ reported in previous studies. This also renders the B\"ootes I
dSph inconsistent with the universal mass profile.

In \citet{walker09b}, velocity dispersions for only 2 M31 dSphs (And II and
IX) were available. We therefore fit NFW and cored density profiles
(equations~\ref{eqn:nfw} and~\ref{eqn:core}) to the velocity dispersions of
the entire dSph population with $L>2\times10^4\lsun$ (ensuring we probe the
same luminosity regime in both the MW and M31), to see how well these
populations can be fit with a single density profile. Both profiles have two
free parameters of interest to fit, the circular velocity of the halo, $V_{\rm
  max}$ and the scale radius $R_S$. To constrain these values, we use a
maximum likelihood fitting routine to determine the most probable values for
these parameters by maximising the likelihood function, $\Lagr$, defined as:

\begin{equation}
\begin{aligned}
\Lagr(\{r_{h,i},\sigma_{v,i},\delta_{\sigma_v,i}\}|V_{\rm max},R_S)=\prod_{i=0}^{N} \frac{1}{\sqrt{2\pi\delta_{\sigma_v,i}^2}}\\{\rm exp}\left[-\frac{(\sigma_{\mathrm{profile}}-\sigma_{v,i})^2}{2\delta_{\sigma_v,i}^2}\right]
\end{aligned}
\label{eqn:maxlike}
\end{equation}

\noindent where $\sigma_{\mathrm{profile}}$ is the velocity dispersion as
predicted by equations~\ref{eqn:nfw} and \ref{eqn:core} for a dSph with
half-light radius $r_{h,i}$; $\sigma_{v,i}$ is the measured velocity
dispersion of the $i^{{\rm th}}$ dSph and $\delta_{\sigma_v,i}$ is the
uncertainty on the measured dispersion. We include only the uncertainty in
$\sigma_v$ in our method, neglecting that of the half-light radius, as the
velocity dispersion parameter has a greater impact on the mass profile, as it
is proportional to the square of $\sigma_v$, depending only linearly on
$r_{\rm half}$..

We show the results of this fit in Fig.~\ref{fig:universal}. The red triangles
represent MW dSphs brighter than $L=2\times10^4\lsun$, while open triangles
represent those fainter than this cut. The blue circles are the M31 dSphs. The
magenta dot-dashed line shows our best fit NFW profile to the whole population
(with $V_{\rm max}=14.7\pm0.5\kms$ and $R_S=876\pm284$~pc), whilst the cyan
dashed line is the best fit core profile (with $V_{\rm max}=14.0\pm0.4\kms$
and $R_S=242\pm124$~pc) where the best fit $V_{\rm max}$ ($R_S$) parameter is
determined by marginalizing the 2D maximum likelihood contours over $R_S$
($V_{\rm max}$). In all cases, the quoted uncertainties are derived by
assuming that the likelihood functions have a Gaussian-like distribution,
allowing us to project the marginalized maximum likelihood contours to the
value at which $2\mathrm{ln}(\Lagr)$ has decreased by the square of the
confidence interval of interest, which in this case is the 1$\sigma$
(i.e. 68\%) confidence interval. Clearly, neither of these mass profiles is a
good fit for many of the Local Group dSphs. This is statistically demonstrated
by the reduced $\chi^2$ values for these fits ($\chi^2=4.1$ for both
profiles). Of the 39 objects, 24 are outliers at $>1\sigma$, with $\sim1/5$ of
the population being outliers at the $3\sigma$ level.

\subsection{Scatter about an average mass profile}
\label{sect:scatter}

From the above analysis, it is clear that the scatter about the best fitting
profiles is significant, and well beyond what we can hope to explain with
measurement uncertainties. But do we really expect that all low luminosity galaxies
should reside in dark matter halos with identical density profiles? The dark
matter subhalos produced in, for example, the Aquarius simulations
\citep{springel08} demonstrate a range of possible values of $V_{\rm max}$ and
$R_{s}$ for these objects. Further, work by e.g. \citet{zolotov12,brooks12}
and \citet{veraciro13} demonstrate that the infall time, host mass and
presence of baryons can all effect the dark matter structures of subhalos. As
such, scatter in density profiles is completely expected, and differences
between the satellite populations of the Milky Way and Andromeda might also be
seen that could tell us about the evolutionary histories of the two systems.

To investigate this, we introduce a mass-scatter term, $\sigma_{V_{\rm max}}$, 
into our maximum likelihood fitting algorithm (equation~\ref{eqn:maxlike})
replacing $\delta_{\sigma_v,i}$ with $\delta_{\mathrm{tot},i}$ which is the combination of the
measured uncertainty in the velocity dispersion measurements and the mass-scatter
term, such that $\delta_{\mathrm{tot},i}=\sqrt{\delta_{\sigma_v,i}^2+\sigma_{V_{\rm max}}^2}$.

\begin{figure*}
  \begin{center}
    \includegraphics[angle=0,width=0.45\hsize]{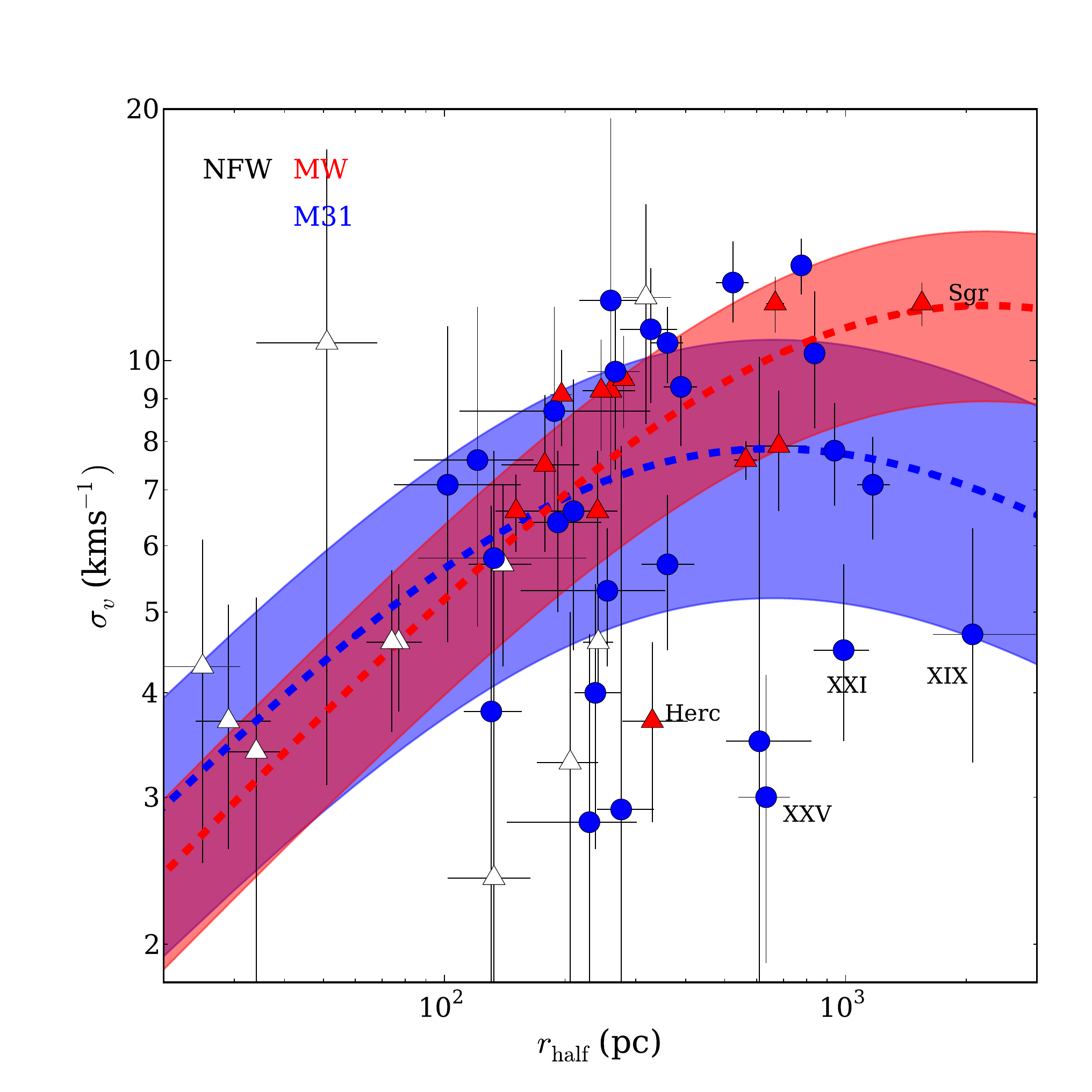}
    \includegraphics[angle=0,width=0.45\hsize]{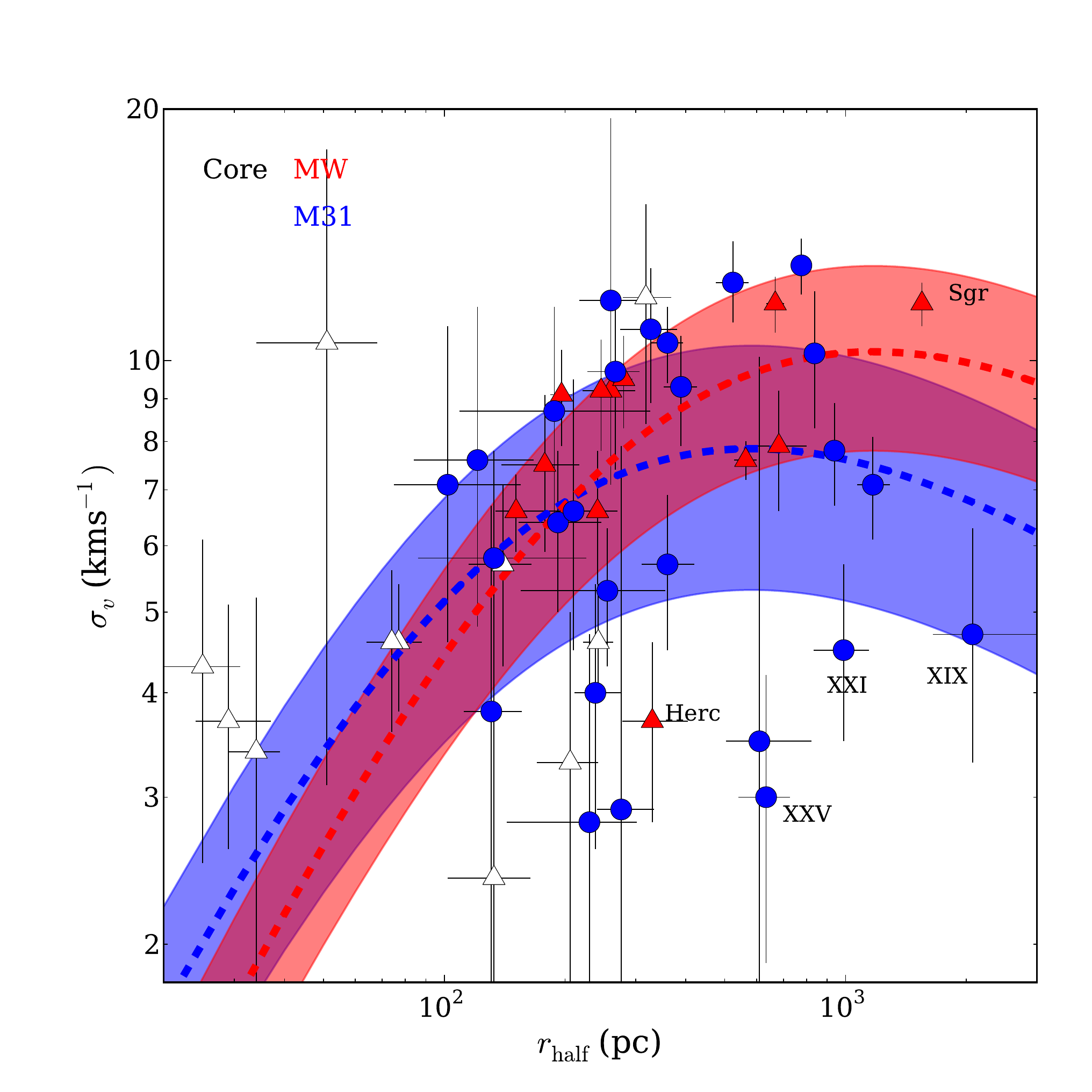}
   \includegraphics[angle=0,width=0.45\hsize]{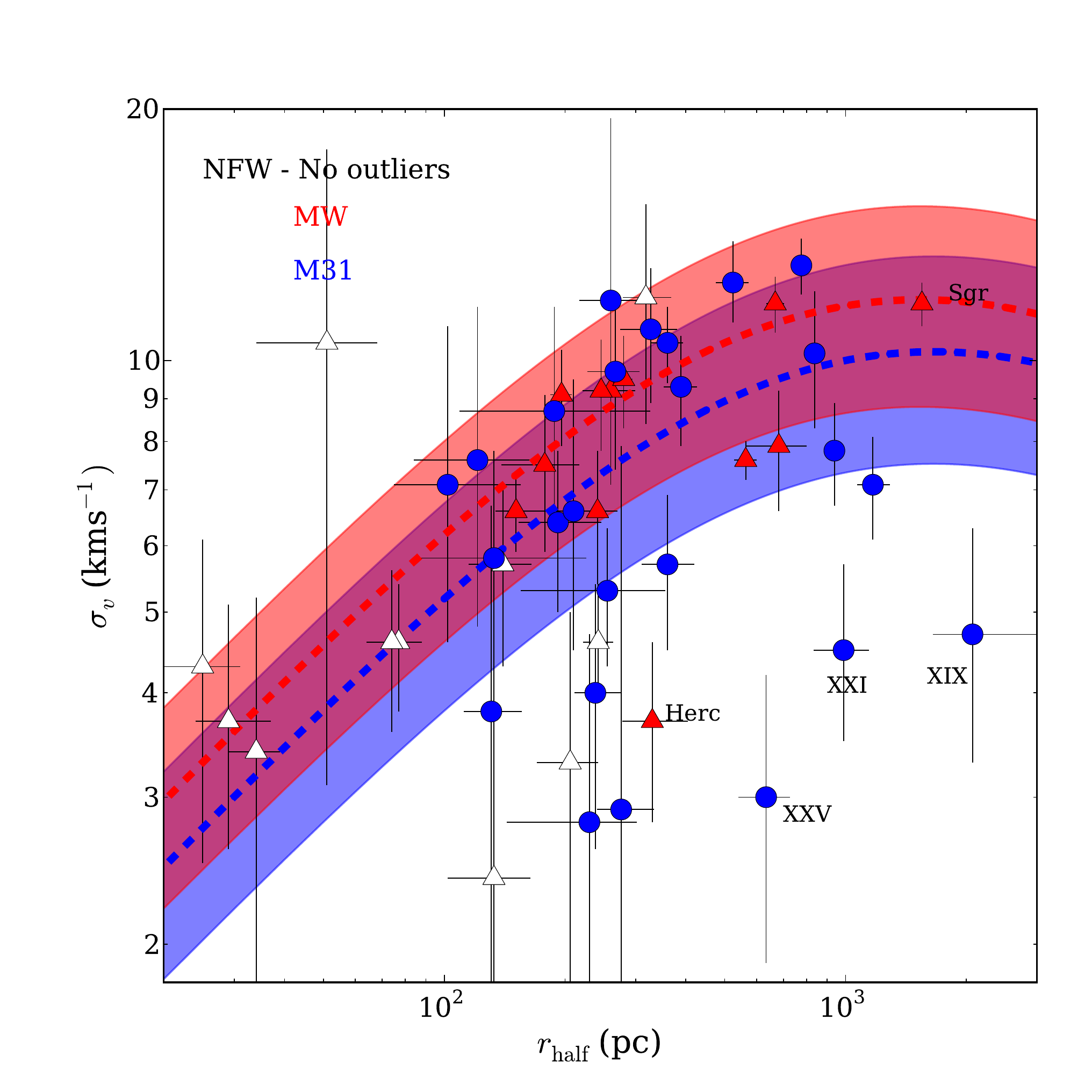}
    \includegraphics[angle=0,width=0.45\hsize]{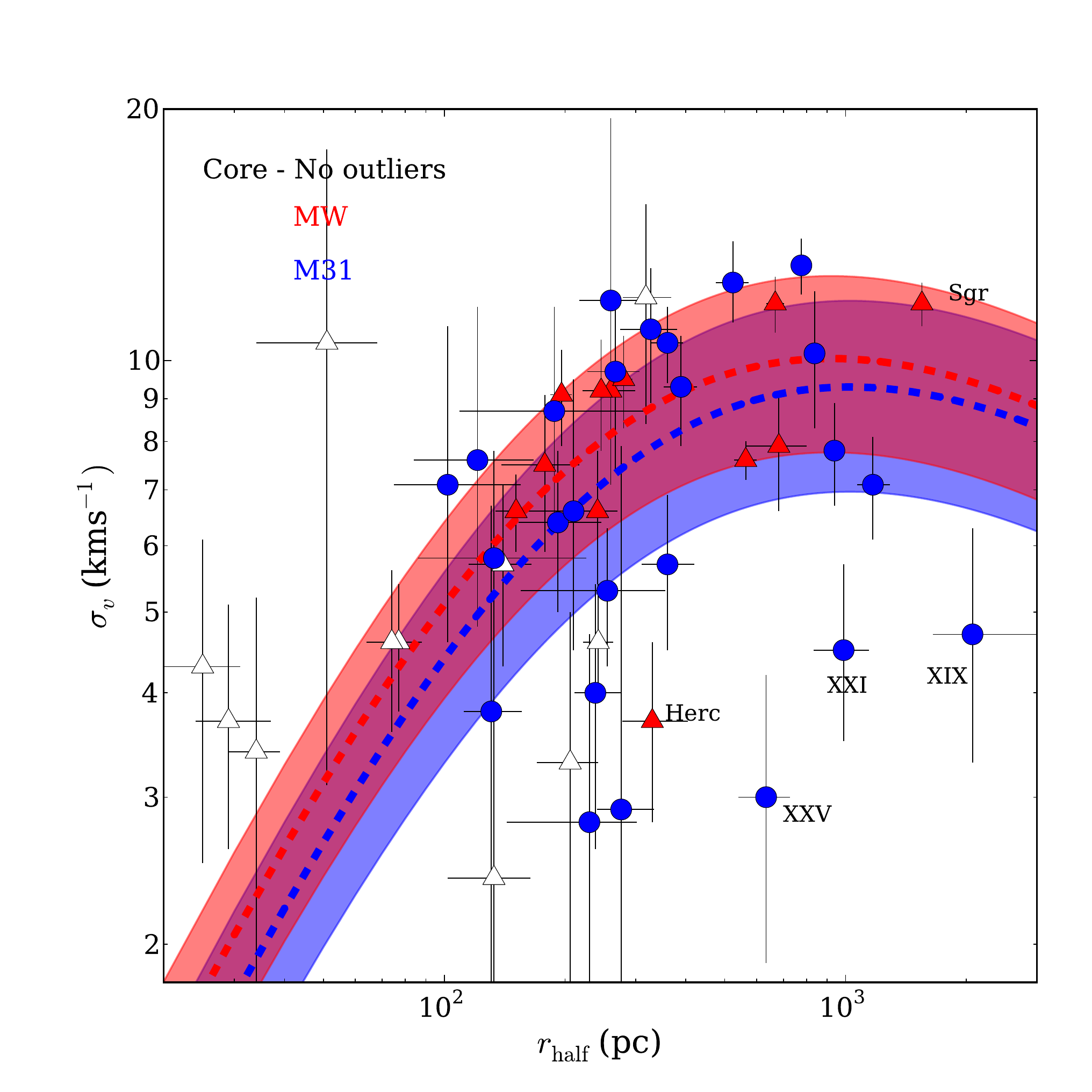}
    \caption{{\bf Top left:} $\sigma_v$ vs. $r_{\rm half}$ for all MW (red
      triangles) and M31 (blue circles) dSph galaxies. Overlaid are the best
      fit NFW profiles for the MW (red shaded region) and M31 (blue shaded
      region). The dashed lines represent the average fit while the shaded
      regions indicate the parameter space allowed by the introduction of the
      scatter term, $\sigma_{V_{\rm max}}$, convolved with the uncertainties
      in the fit parameters. At large $r_{\rm half}$ (higher mass) the
      profiles of these populations begin to diverge, with the M31 fit turning
      over at $r_{\rm half}\sim600\pc$ while the MW profile continues to
      rise. {\bf Top right:} As top left, but with the best fit cored density
      profiles overlaid. Again, the M31 profile is seen to turn over before
      that of the MW profile. {\bf
        Bottom left and right: } As top panels, but now the best fit profiles for NFW
      (left) and core (right) are determined after excluding And XIX, XXI and
      XXV. The removal of these objects results in best fit cored mass profile
      parameters that agree extremely well for MW and M31 dSphs.}
  \label{fig:fitsall}
  \end{center}
\end{figure*}

If the mass profiles of the Andromeda and Milky Way dSphs are truly similar
within their inner regions, our algorithm should find best fit values of
$V_{\rm max}$, $R_S$ and $\sigma_{V_{\rm max}}$ that are broadly consistent
when fitting the two populations separately and as a whole. In
Figure~\ref{fig:allML} we show the likelihood contours for these
parameters. In the top three panels we overlay the $1\sigma$ and $2\sigma$
confidence interval contours (again, defined as the region of parameter space
where $2\mathrm{ln}(\Lagr)$ decreases by the square of the confidence interval
in question) for the NFW fits to the Milky Way (red dashed), Andromeda (blue
dot-dashed) and the full sample (black solid) for our 3 free parameters
(marginalized over the 3rd parameter not displayed in each 2D subplot), with
solid points representing the best fit values in each case. In the second row
of subplots, we show the one dimensional marginalised relative likelihood functions for
$V_{\rm max}$, $R_S$ and $\sigma_{V_{\rm max}}$ for each of these fits. The
lower 6 panels show the same, but for the cored fits. In the NFW case, while
the best fit values for $R_S$ in each case seem dramatically different for the
MW and M31 at first glance with \rsnmw\ and \rsna\ respectively, their
uncertainties are such that they agree within $1\sigma$. The amount of scatter
in mass at a given radius is also very similar, with \signmw\ and \signa\
respectively for the MW and M31. The preferred values for $V_{\rm max}$
(\vmaxnmw\ and \vmaxna\ respectively), however, are marginally less
consistent, with M31 preferring a lower value of $V_{\rm max}$ (and hence,
lower masses) than the MW system. For the core profiles, we get a similar
result, with the values for $R_S$ (\rscmw\ and \rsca), and $\sigma_{V_{\rm
    max}}$ (\sigcmw and \sigca) for the MW and M31 agreeing within $1\sigma$,
and marginally inconsistent values of $V_{\rm max}$ (\vmaxcmw\ and \vmaxca).

In the top two panels of Fig.~\ref{fig:fitsall}, we overplot the best fit relations from this
analysis in the $r_{\rm half}-\sigma_v$ plane. In the left panel, we show the
best fit NFW profile for the MW (red line) and M31 (blue line) dSphs, with the
best fit core profiles in the right panel. The shaded regions represent the
scatter we derived convolved with the $1\sigma$ uncertainties for $V_{\rm max}$ and
$\sigma_{V_{\rm max}}$. Both the MW NFW and Core profiles provide an excellent
fit to all the dSphs barring the Hercules dSph. However, as it is likely to be
highly tidally disturbed \citep{aden09,martin10}, this is not too
surprising. For the M31 fits, we see 3 systems that are outliers at the
$\sim2-3\sigma$ level: And VI, VII and XXV. And XXV has previously been
identified as an unusually low mass system in C13, so this inconsistency is
perhaps not unexpected. However, And VI and VII are thought to represent
fairly typical satellites, with velocity dispersions similar to their MW
counterpart Fornax, which has a comparable half-light radius to these two
objects.

\begin{figure*}
  \begin{center}
     \includegraphics[angle=0,width=0.3\hsize]{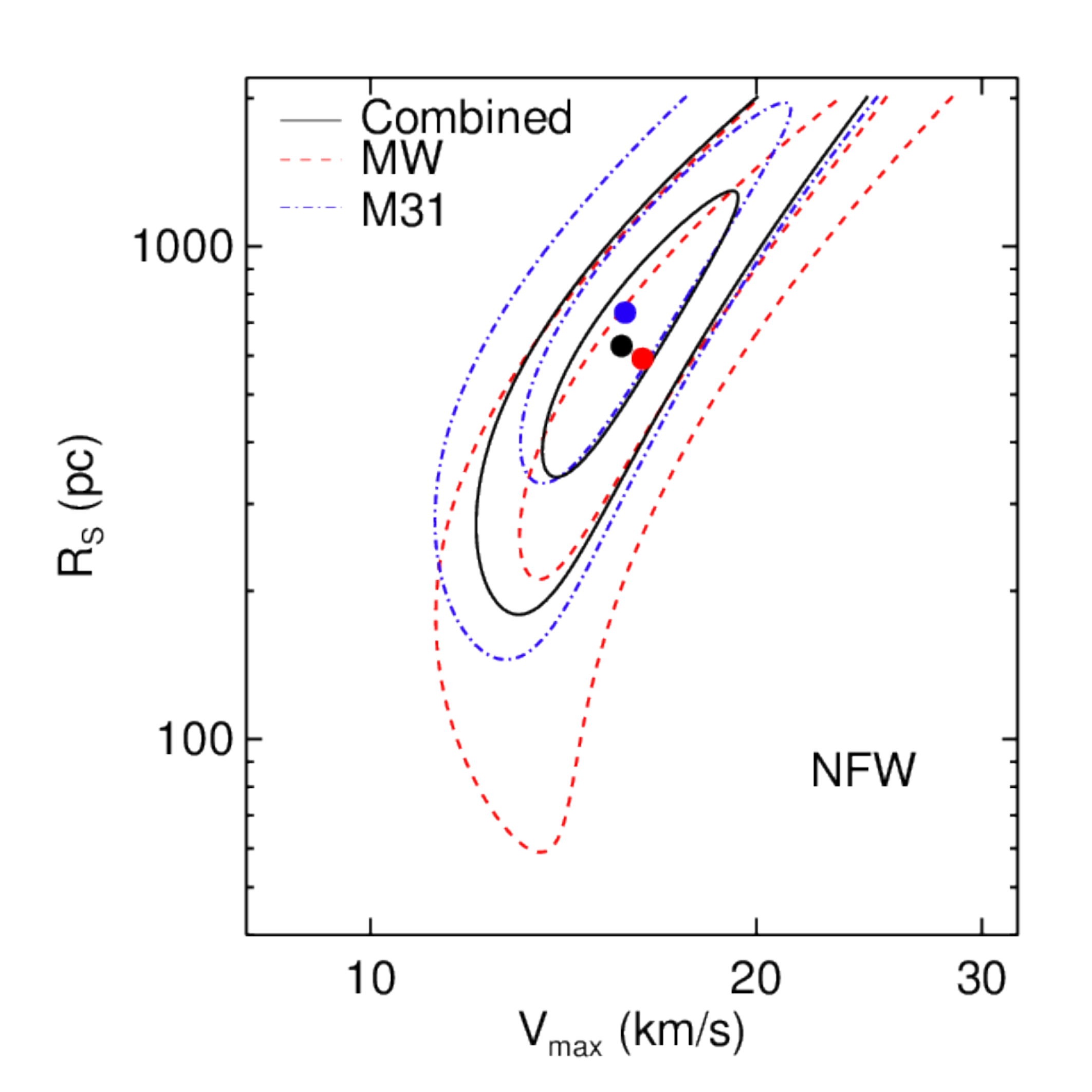}
    \includegraphics[angle=0,width=0.3\hsize]{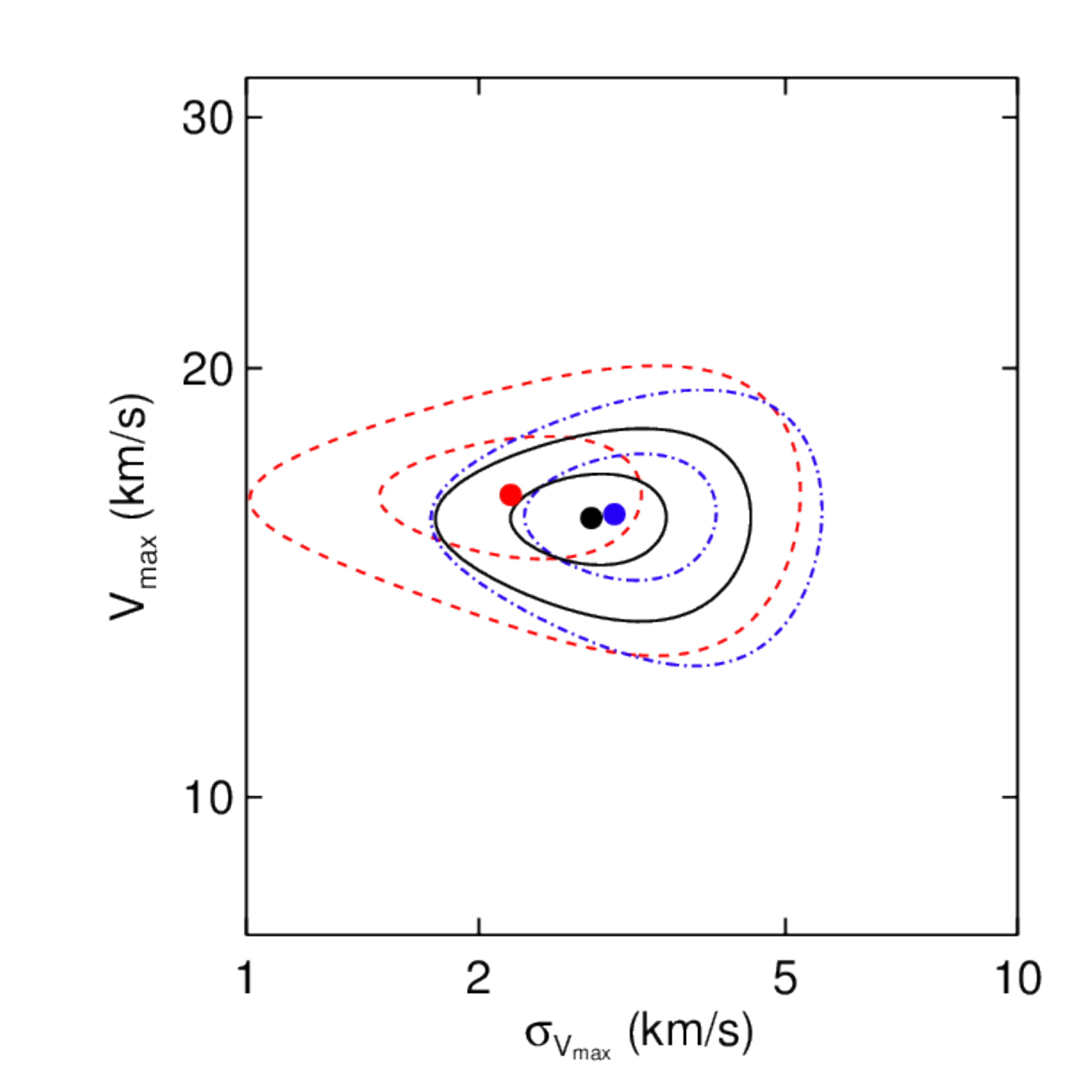}
     \includegraphics[angle=0,width=0.3\hsize]{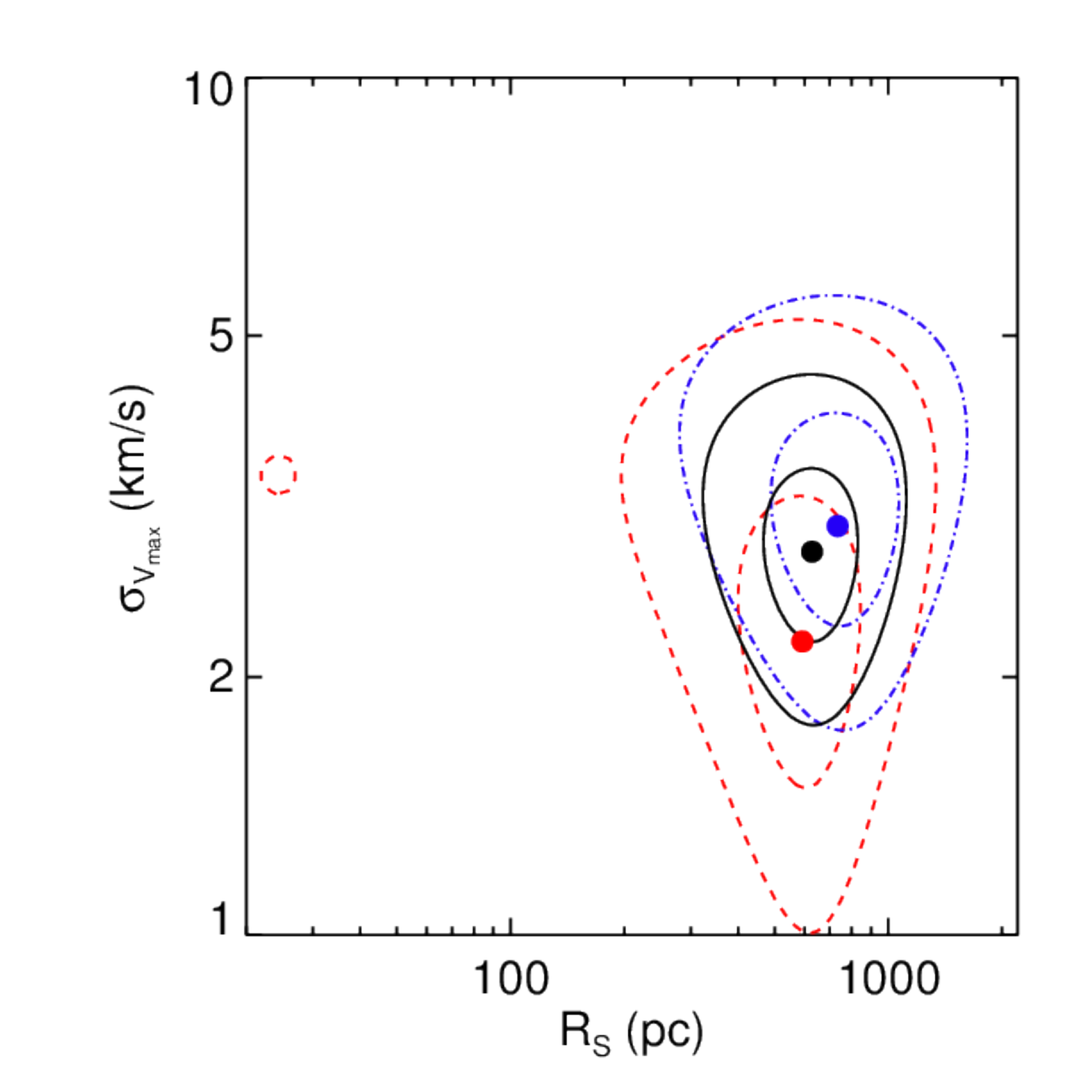}
    \includegraphics[angle=0,width=0.3\hsize]{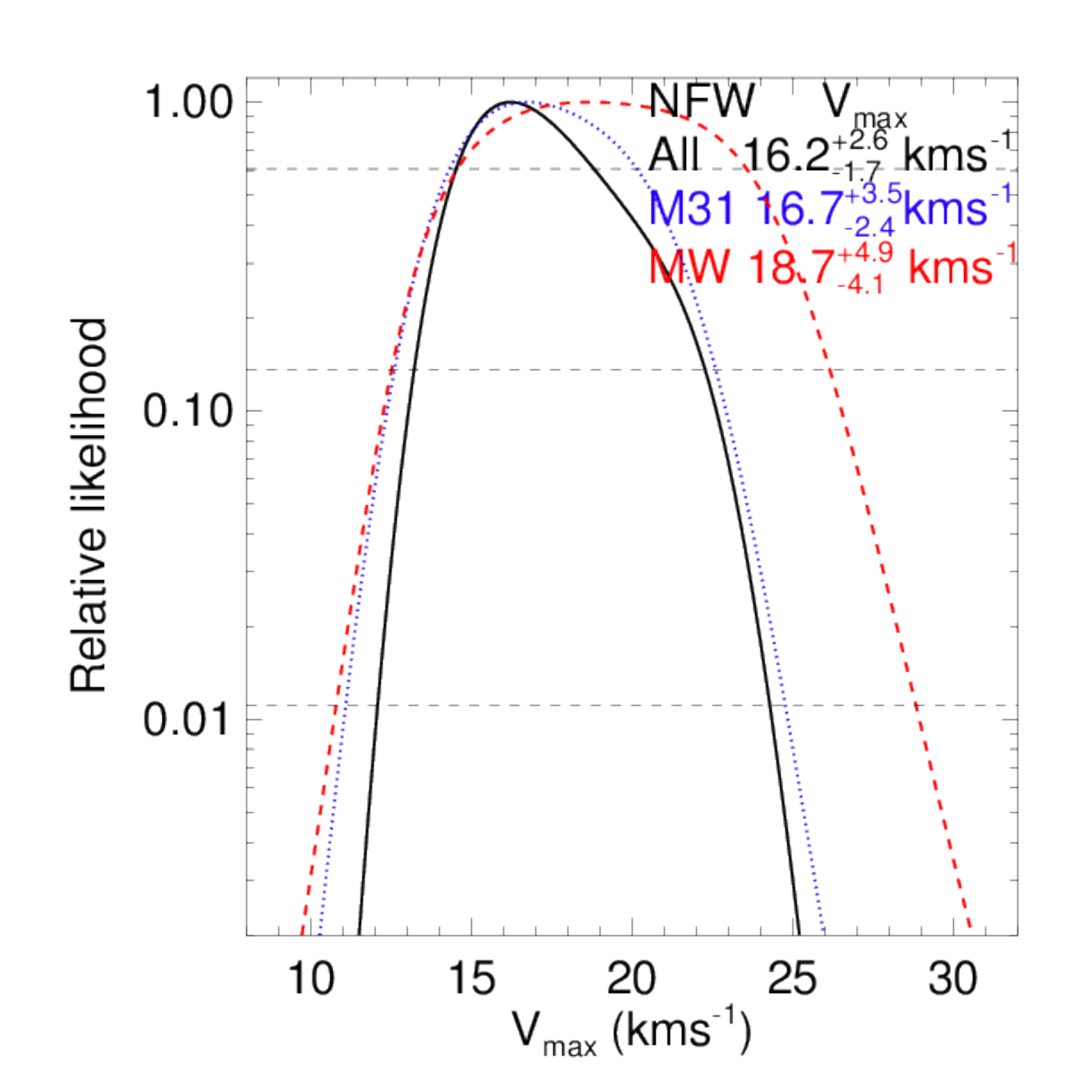}
    \includegraphics[angle=0,width=0.3\hsize]{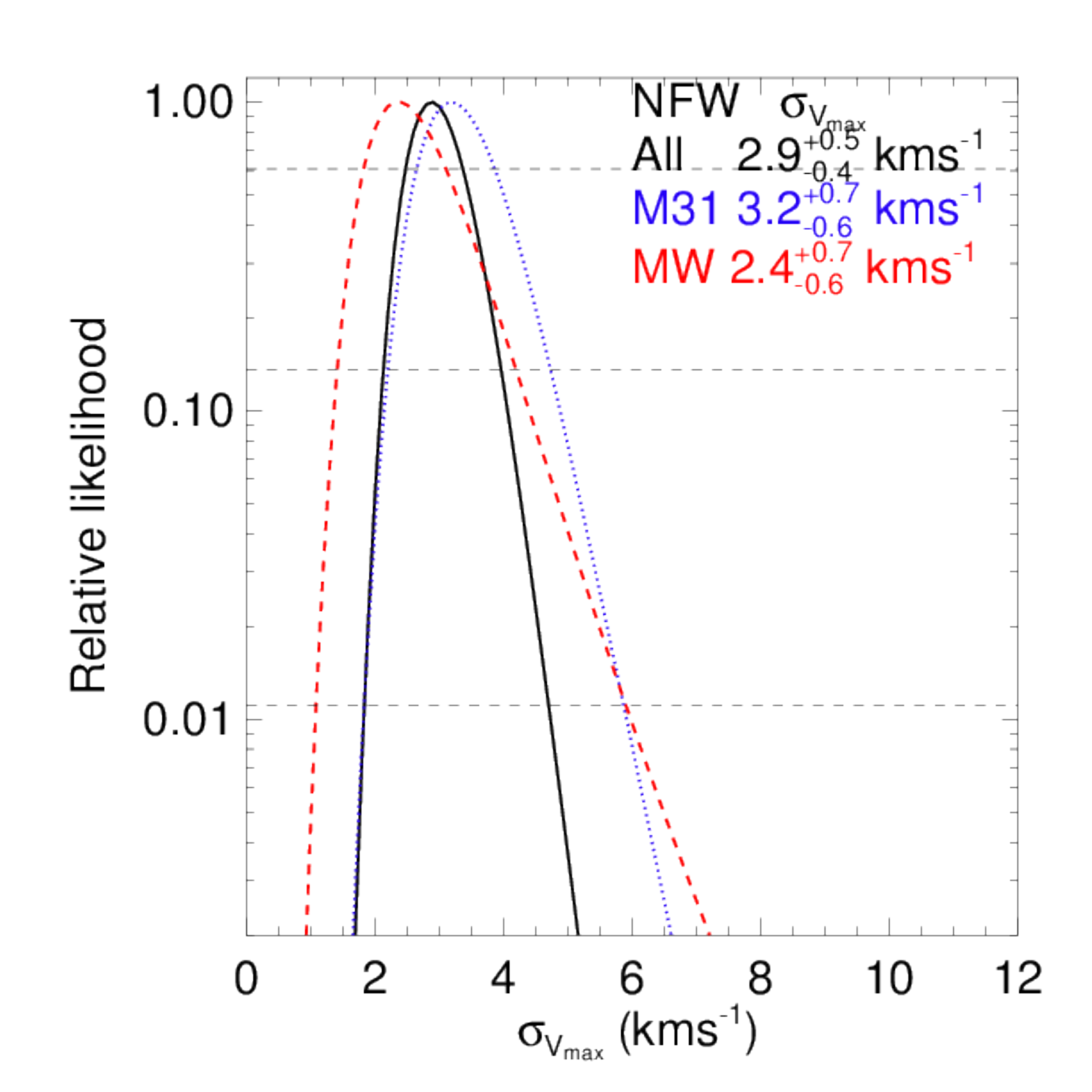}
     \includegraphics[angle=0,width=0.3\hsize]{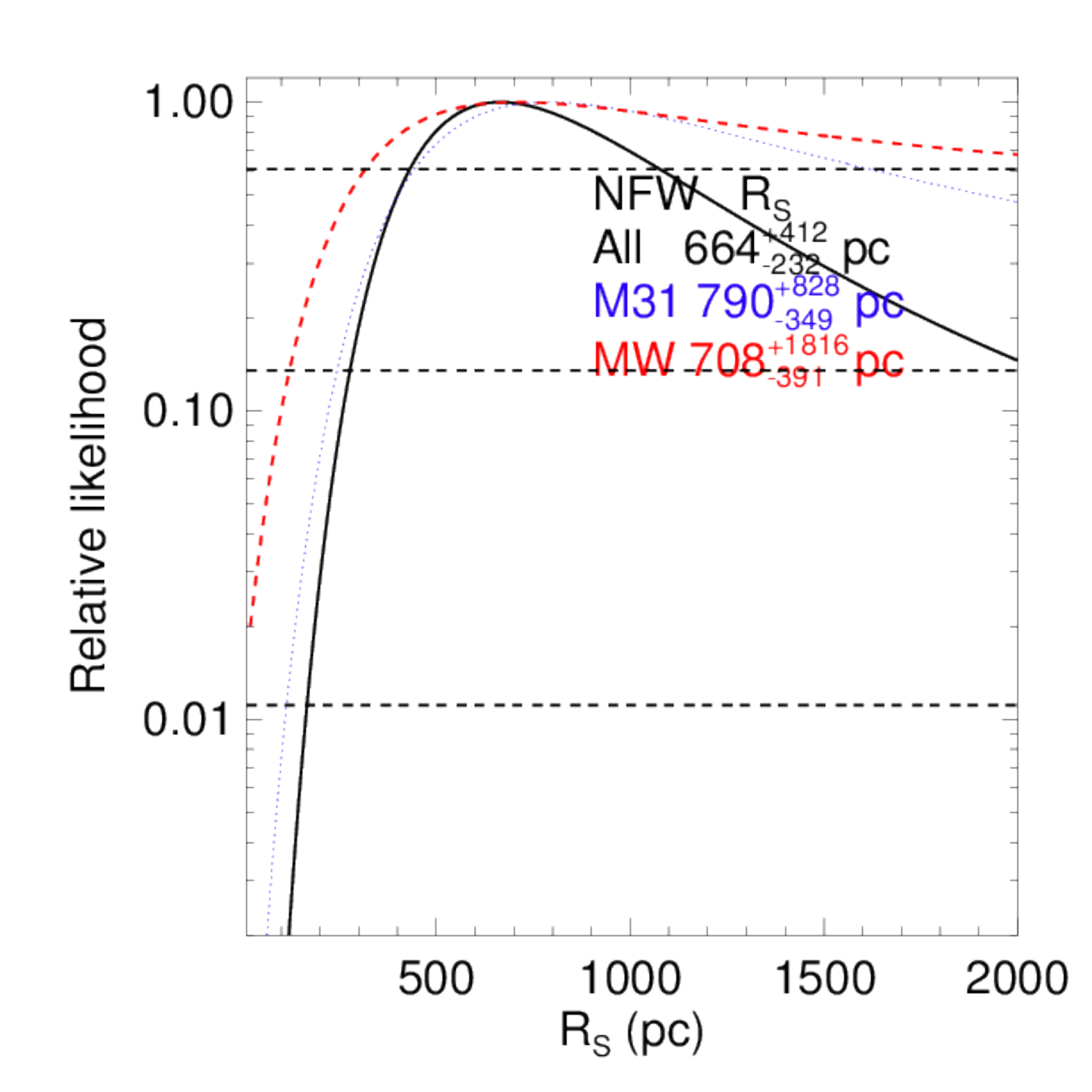}
      \includegraphics[angle=0,width=0.3\hsize]{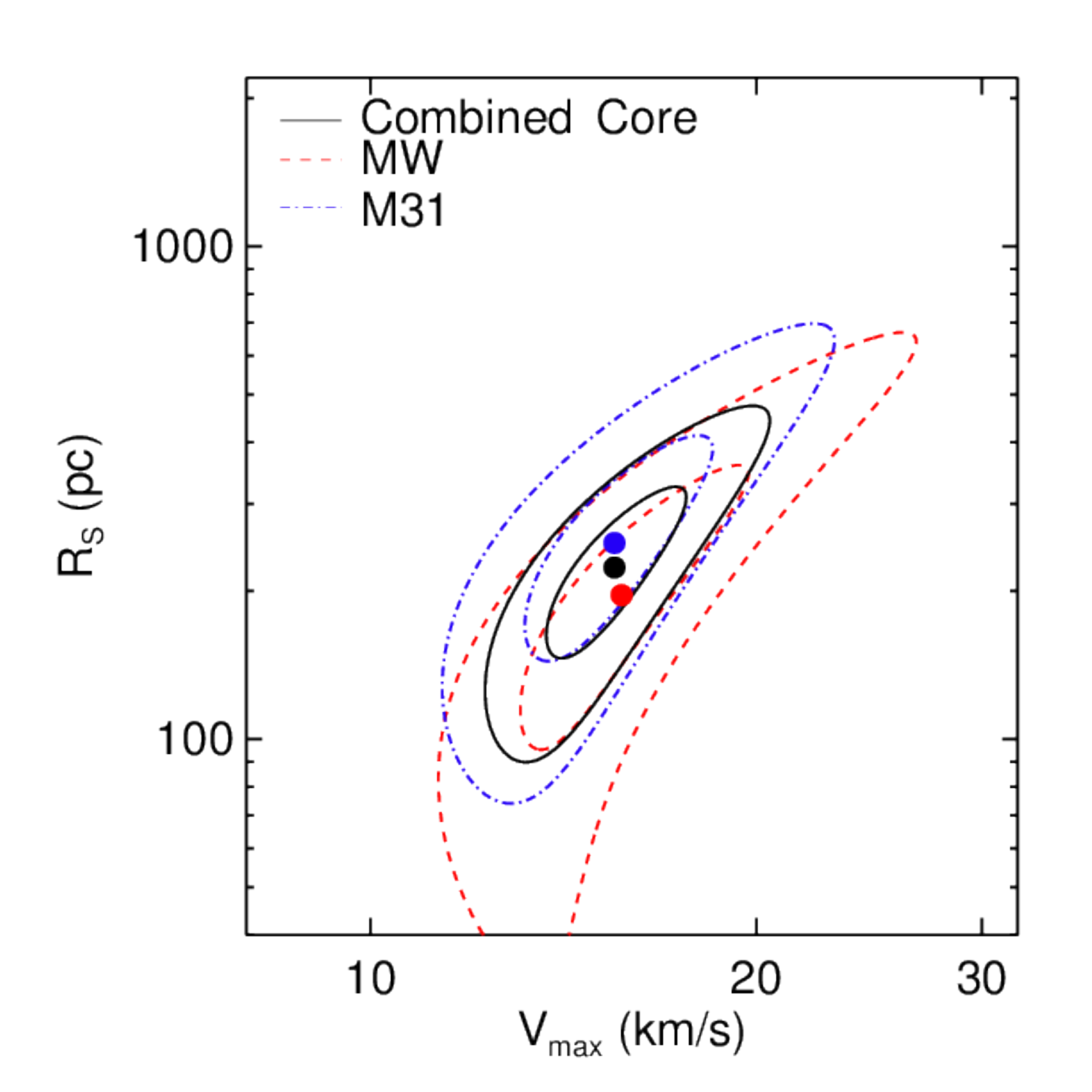}
    \includegraphics[angle=0,width=0.3\hsize]{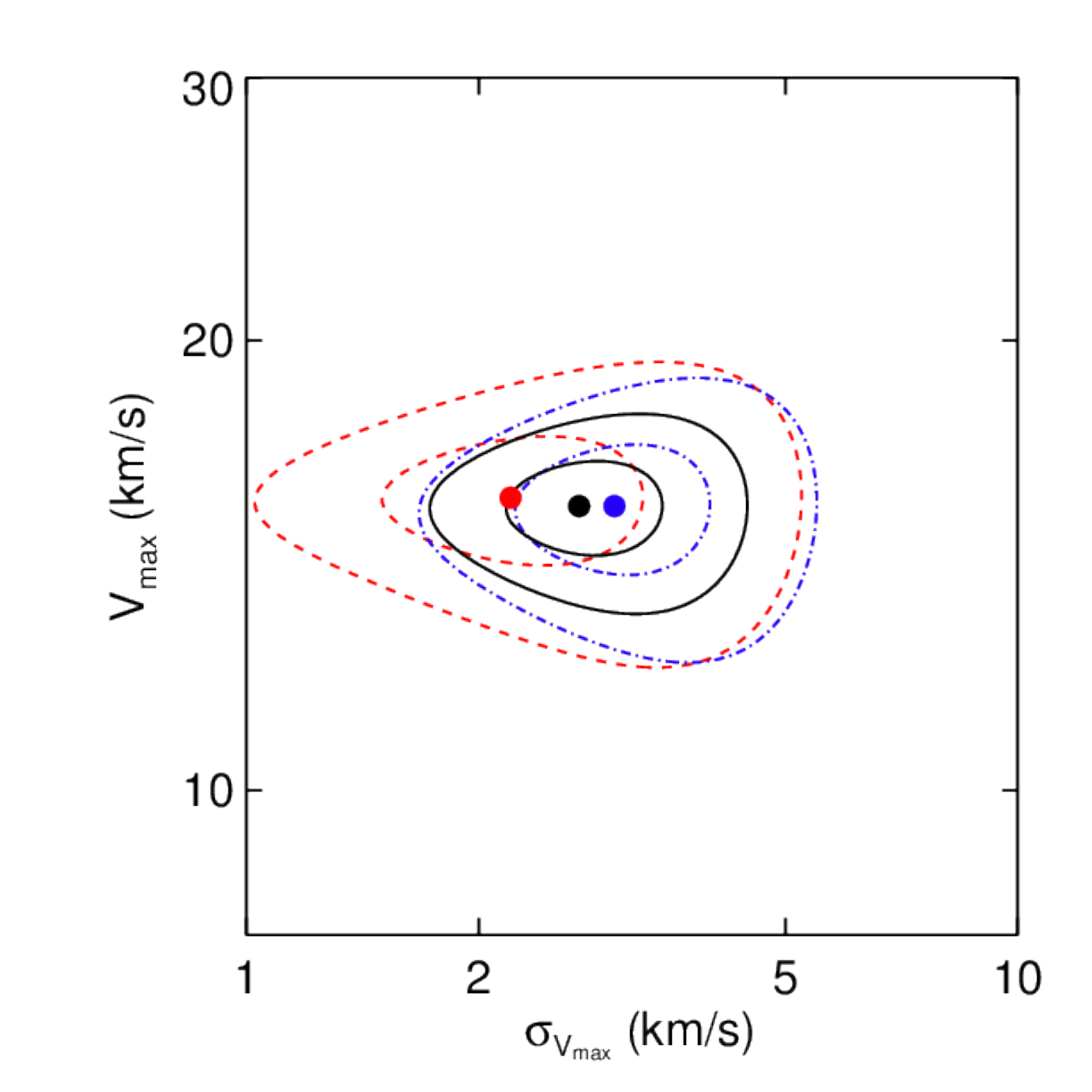}
     \includegraphics[angle=0,width=0.3\hsize]{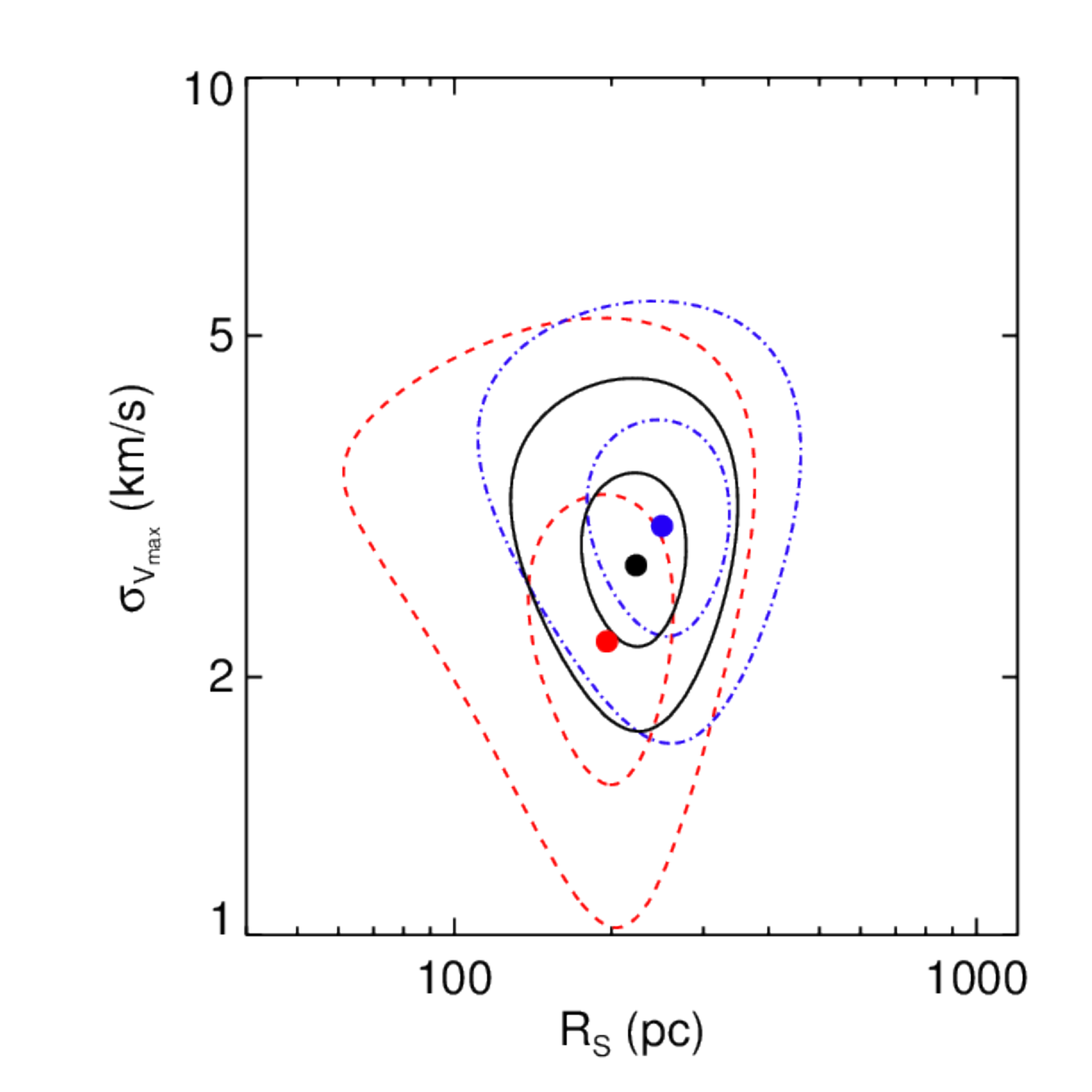}
     \includegraphics[angle=0,width=0.3\hsize]{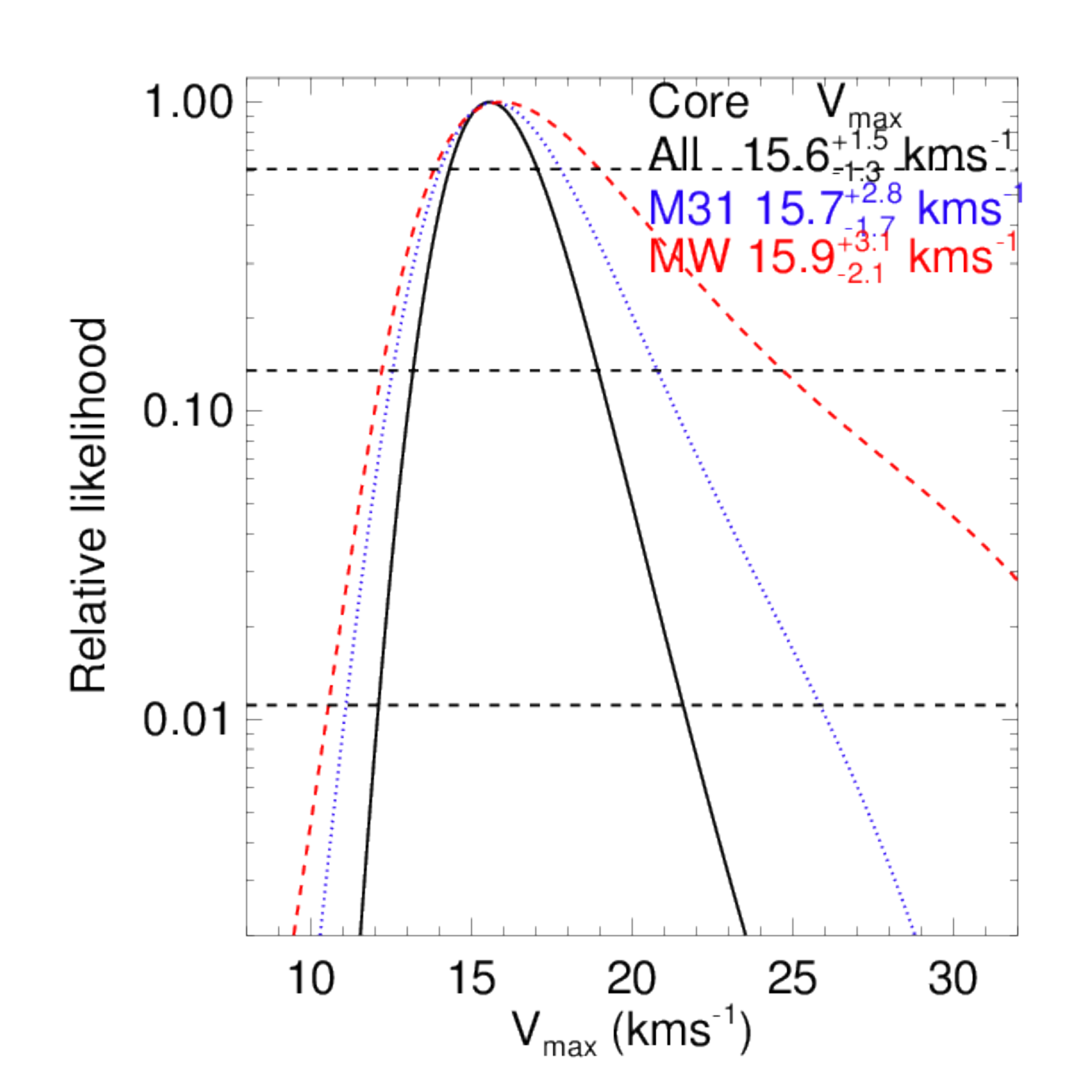}
    \includegraphics[angle=0,width=0.3\hsize]{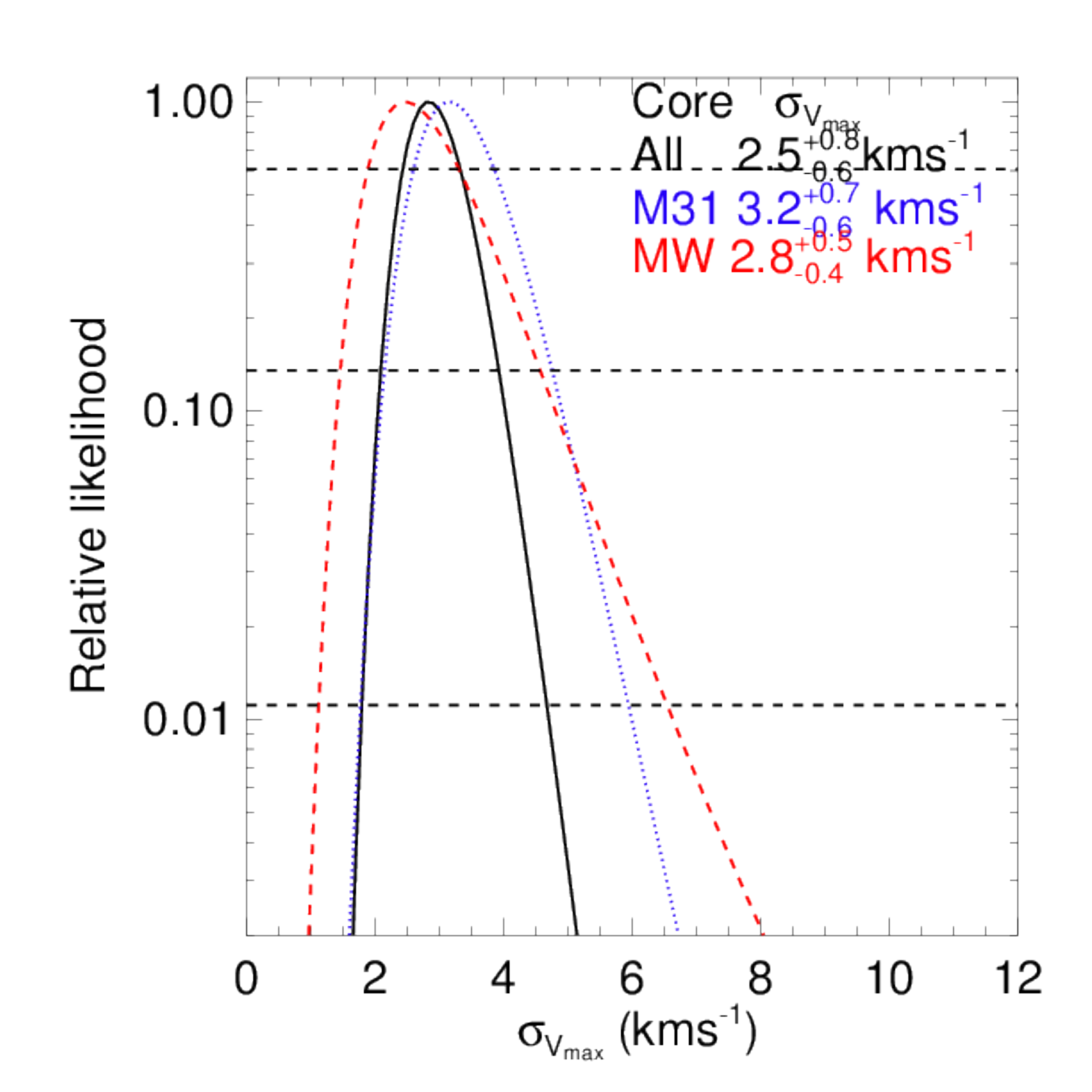}
     \includegraphics[angle=0,width=0.3\hsize]{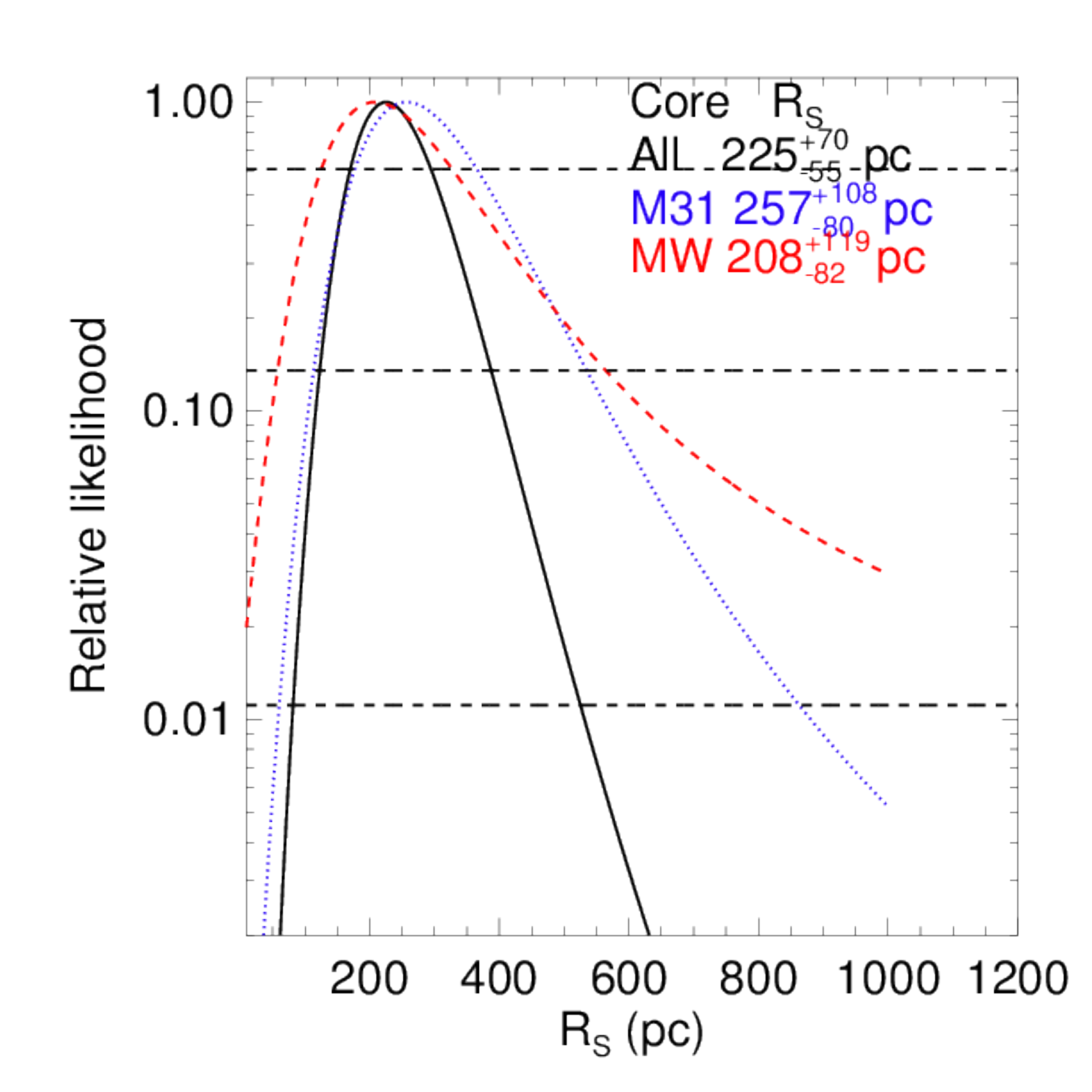}   \caption{As Fig.~\ref{fig:allML}, but with three significant low mass
      outliers, And XIX, XXI and XXV, omitted from the fits. The removal of
      these objects results in best fit NFW mass profile parameters that agree
      extremely well for MW and M31 dSphs. }
  \label{fig:outML}
  \end{center}
\end{figure*}

The differences in the preferred $V_{\rm max}$ has a striking visual effect on
the resulting best fit profiles for the MW and M31. While in both the NFW and
Core case, the relations for MW and M31 populations track each other well at
the smaller radius (lower mass) end (albeit with greater scatter in M31), at
larger radii there appears to be a divergence between the two systems. Both
the NFW and core profiles for M31 begin to turnover at $\sim600$~pc, while the
MW profile continues to rise (turning over at $\sim1200$~pc in the cored
case). This turnover radius is interesting, as there are only 3 MW dSphs with
half-light radii $\gta600$~pc, one of which is the tidally disrupting
Sagittarius (Sgr) and thus is excluded from our fits, the other two being
Fornax and Sextans. In M31, there are 7 galaxies (And I, II, VII, XIX, XXI,
XXIII and XXV), 3 of which (And XIX, XXI and XXV) are curiously very low mass
for their size. In C13, they were measured to be $3\sigma$ outliers to
the best-fit mass profiles of \citet{walker09b} and, as can be seen in
Fig~\ref{fig:fitsall}, they are significant outliers to the
best fit MW relation ($2,2.5$ and $3\sigma$ respectively). In addition,
despite having very similar half-light radii, Sgr ($r_{\rm
  half}=1550\pm50$~pc) and And XIX ($r_{\rm half}=2072^{+1092}_{-422}$~pc)
have very different velocity dispersions ($\sigma_v=11.7\pm0.7\kms$ and
$\sigma_v=4.7^{+1.6}_{-1.4}\kms$).

\begin{figure}
  \begin{center}
      \includegraphics[angle=0,width=0.9\hsize]{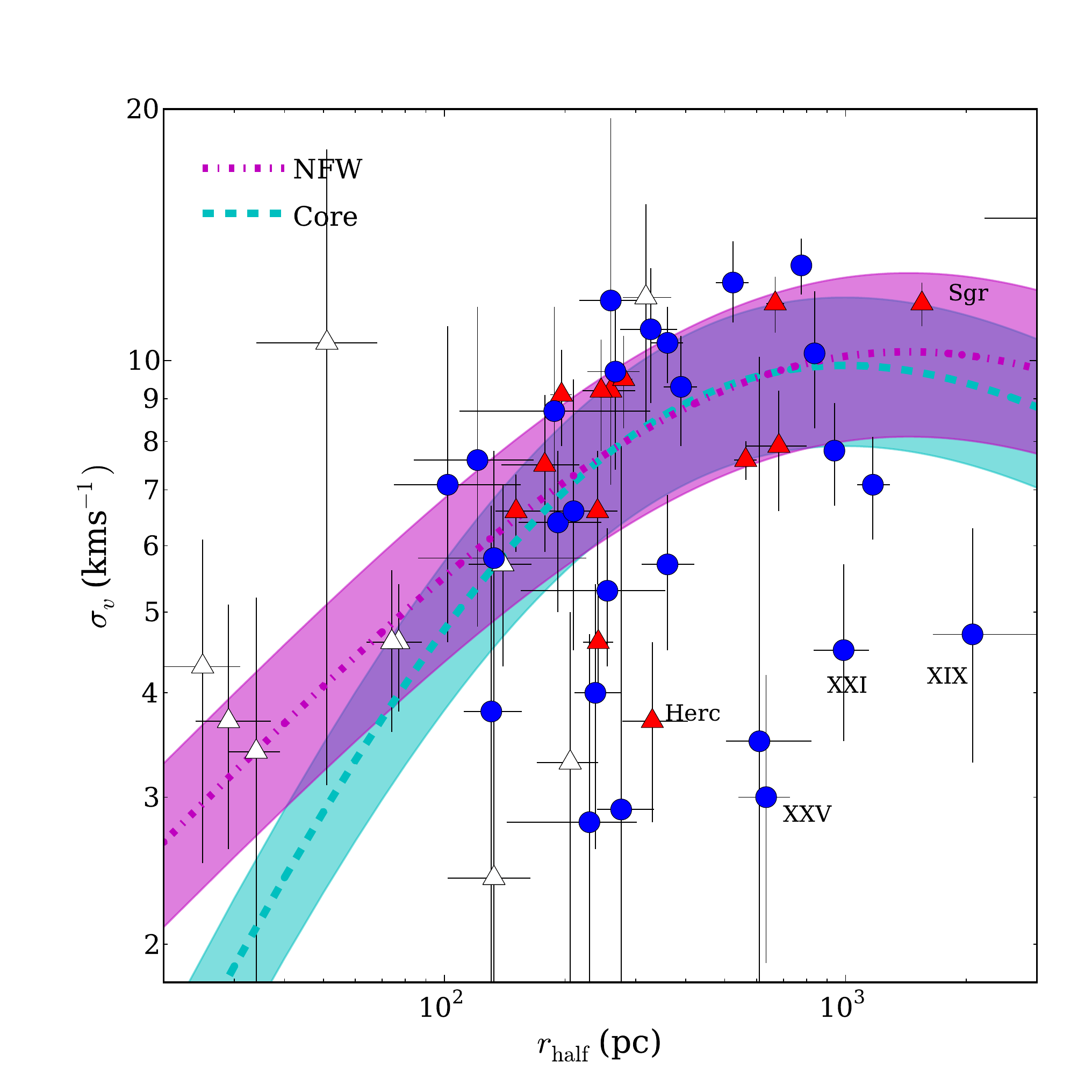}
      \includegraphics[angle=0,width=0.9\hsize]{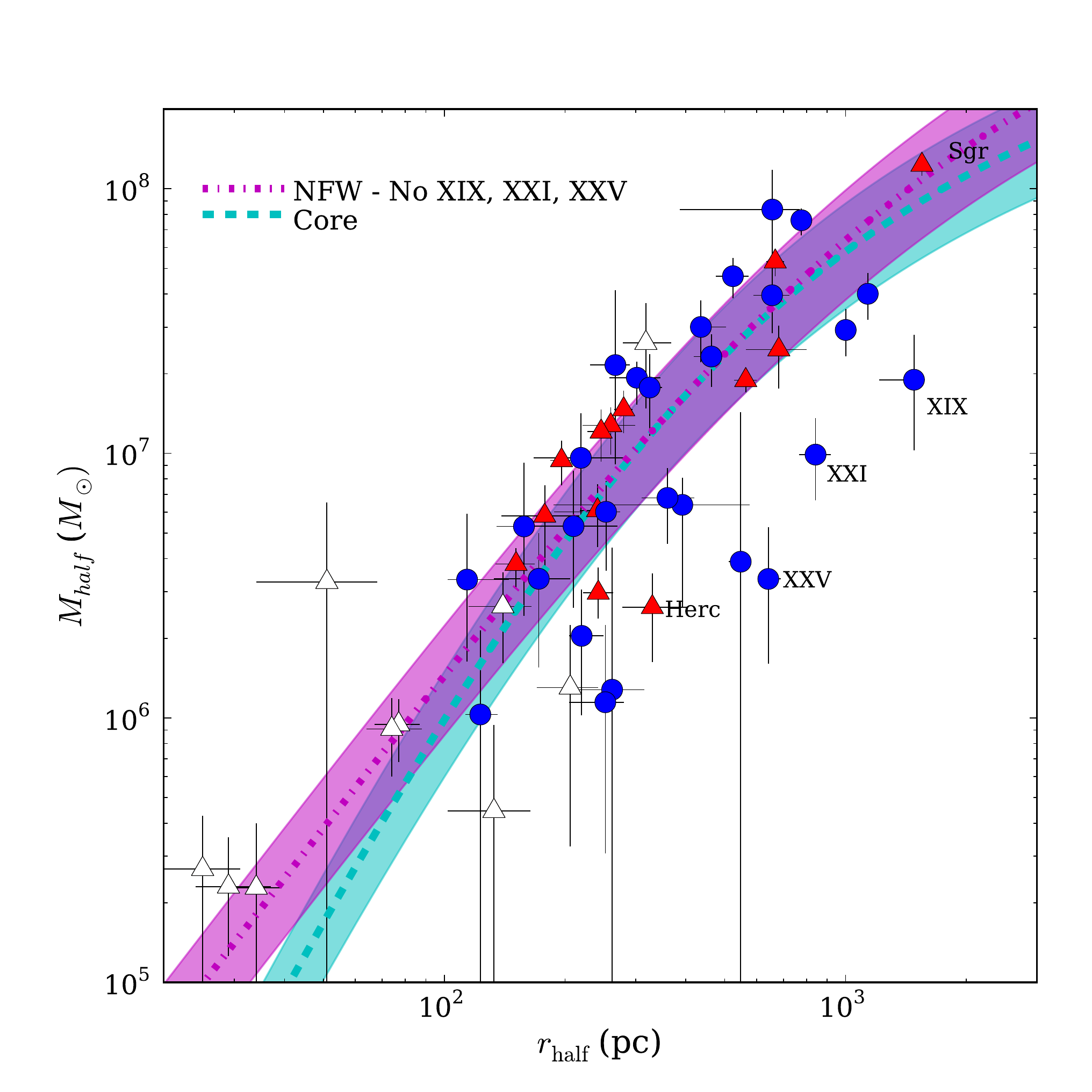} \caption{ The best
        fit NFW (magenta) and core (cyan) relations to the whole Local Group
        population as seen in the $r_{\rm half}-\sigma_v$ (top) and $r_{\rm
          half}-M_{\rm half}$ (bottom) planes.  We see that the velocity
        dispersions and masses for all the dSphs, barring the excluded
        outliers and 3 further objects (discussed further in the text) agree
        with the fits to the whole Local Group population within their
        uncertainties. }
  \label{fig:fitLG}
  \end{center}
\end{figure}

To see if it is these outliers driving the differences between the MW and M31
relations, we repeat the same fits performed above, but without all systems
that were found to lie at or greater than $3\sigma$ from the best fit profiles
of \citet{walker09b}. This included the three M31 outliers, and additionally
the MW dSphs, Hercules and CVn I, which are also outliers to the
\citet{walker09b} relation. In Fig.~\ref{fig:outML} we show the same contours
as in Fig.~\ref{fig:allML} for NFW and core profile fits, only this time with
these outliers omitted. The exclusion of Hercules and CVn I from the MW fits
has only a slight effect on these profiles, but removing the low mass M31
outliers is more substantial. The agreement between $R_s$, $V_{\rm max}$ and
$\sigma_{V_{\rm max}}$ is significantly better (as shown in
Table~\ref{tab:profiles}). In the lower two panels of Fig.~\ref{fig:fitsall},
we show these best fit profiles in the $r_{\rm half}-\sigma_v$ plane. Barring
the 5 excluded dSphs and the M33 satellite, And XXII \citep{chapman12}, all
the Local Group objects have velocity dispersions that are well described by
the M31 and MW relations. The best fit profiles to the whole Local Group
system are shown in Fig.~\ref{fig:fitLG}, and have preferred values for the
NFW (core) parameters of \rsno\ (\rsco), \vmaxno\ (\vmaxco) and \signo\
(\sigco). Therefore, whilst dSph galaxies do not live within dark matter halos
with identical density profiles, the vast majority do inhabit statistically
similar halos with a well defined mass range at any given radius.

In Fig.~\ref{fig:fitLG} we show the mass within the half-light radii of the
dSphs (calculated using the \citealt{walker09b} mass estimator, where $M_{\rm
  half}=580r_{\rm half}\sigma_v^2$, tabulated in Table~\ref{tab:ml}) with the best fit NFW and Core relations
when excluding the outliers. At all radii, the total scatter is less than half
a magnitude in mass. For example, at $r_{\rm half}=10\pc,100\pc$ and 1000 pc,
the average masses from the cored profile are $M_{\rm
  half}=2.3\times10^4\msun,1.1\times10^6\msun$ and $5.4\times10^7\msun$, and
the scatter (i.e. half the distance outlined by the shaded band) is $M_{\rm
  scatter}=1.2\times10^4\msun,0.6\times10^6\msun$ and $2.5\times10^7\msun$,
which is $\sim50\%$ of the average mass in each case. The numbers for the best
fit NFW profile are almost identical.

Our decision to exclude And XIX, XXI and XXV was based on their designation as
significant ($>3\sigma$) low mass outliers in C13, and to the derived MW
profile. There are several other potentially low mass systems that were
identified in C13 (namely And XXII) and T12 (And XIV, And XV and XXII
also), which we did not exclude, simply because their associated uncertainties
place them much closer to the regime of expected mass from the MW system. If
they too were shown to be truly low mass with subsequent observations, this
would imply that the M31 dwarf spheroidal systems do have greater scatter
towards lower masses in their mass profiles compared with the MW.

\begin{deluxetable*}{lccccccccc}
\tabletypesize{\footnotesize}
\tablecolumns{10} 
\tablewidth{0pt}
\tablecaption{Best fit parameters from mass profile fits to MW and M31 dSph
  data using NFW and cored profiles.}
\tablehead{
\colhead{Model} &   &  \colhead{Full} &   &  & \colhead{M31} & & &
\colhead{MW} & \\ & \colhead{$V_{\rm max}$}
& \colhead{$R_S$ } & \colhead{$\sigma_{V_{\rm max}} $} & \colhead{$V_{\rm max}$}
& \colhead{$R_S$ } & \colhead{$\sigma_{V_{\rm max}} $} & \colhead{$V_{\rm max}$}
& \colhead{$R_S$ } & \colhead{$\sigma_{V_{\rm max}}$}\\ &
\colhead{($\kms$)} & \colhead{(pc)} & \colhead{($\kms$)} & \colhead{($\kms$)} & \colhead{(pc)} & \colhead{($\kms$)} &\colhead{($\kms$)} & \colhead{(pc)} & \colhead{($\kms$)} }
\startdata
NFW & \vmn & \rn & \sn & \vmna & \rna & \sna & \vmnmw & \rnmw & \snmw\\
NFW (minus outliers) &\vmno & \rno & \sno & \vmnoa & \rnoa & \snoa & \vmnomw & \rnomw & \snomw\\
Core &\vmc & \rc & \scall & \vmca & \rca & \sca & \vmcmw & \rcmw & \scmw\\
Core (minus outliers) &\vmco & \rco & \sco & \vmcoa & \rcoa & \scoa & \vmcomw & \rcomw & \scomw\
\enddata
\label{tab:profiles}
\end{deluxetable*}

\subsection{Comparing  the observational scatter to simulations}
\label{sect:obssims}

Briefly, we compare the best fit values of $V_{\rm max}$ and the scatter in
this term with recent cosmological and semi-analytical models to deduce
whether the values we statistically obtain for the Local Group dSphs compare
favorably with our theoretical understanding of galaxy formation and
evolution. 

If we naively compare to dark-matter only simulations, such as the subhalos in
the Aquarius simulations \citep{springel08} of 6 MW-mass dark matter halos, we
find that our measured values of $V_{\rm max}$ are lower than would be
expected. The same discrepancy was pointed out by \citet{kolchin12}. While the
MW dSphs have $12\lta V_{\rm max}\lta25\kms$, they found at least 10 subhalos
in each Aquarius host with $V_{\rm max}>25\kms$. This discrepancy is referred
to as the `Too Big To Fail' problem, and would seem to persist when
including M31 dSphs.

If we instead compare with models where baryons are taken into account, do we
still see such inconsistencies?  In \citet{rashkov12}, dark matter subhalos
from the high resolution Via Lactea II simulations are populated with baryons
at the time of infall into their host halo by dynamically tagging dark matter
particles as stars. These systems are then traced until $z=0$, where their
final properties are compared to observations. These simulations are able to
reproduce many observed properties of MW dSphs (such as velocity dispersions,
sizes, metalicities, number count etc.), and that the present day values of
$V_{\rm max}$ for the 10 most luminous subhalos are more compatible with
observations, having $10<V_{\rm max}<40\kms$ ($\sim50\%$ of which are less
than $20\kms$). Our average $V_{\rm max}$ plus scatter term gives a
statistical range for the $V_{\rm max}$ of the Local Group dSphs of
$\sim12-22\kms$. So while the bulk of their sample is consistent, there remain
too many high mass dSph satellites to be fully consistent. We can also
  compare our calculated values of $R_{S}$ with those of the \citet{rashkov12}
  simulations. The range of $R_{\rm max}$ (which is the radius at which the
  circular velocity of the halo is at a maximum, i.e., equal to $V_{\rm max}$)
  for their 10 most luminous subhalos ranges from $\sim790-5400$ pc (assuming
  an NFW profile). According to \citet{penarrubia08a}, $R_{\rm max}\sim2R_S$,
  so this corresponds to scale radii for the \citet{rashkov12} halos of
  $395\pc\lta R_s\lta2700$~pc which is consistent with the scale radius of
  \rsno\ that we find for our combined NFW profile (with outliers excluded),
  suggesting that these subhalos are similarly dense to the Local Group
  dSphs.

\citet{bovill11a} model satellites within the Local Volume from reionization
until today, tracing the merger histories and tidal interactions of these
objects as they merge to form more massive galaxies. As with the
\citet{rashkov12} study, they are able to reproduce many of the observed
properties of MW and M31 dSphs. For satellites with similar luminosities to
those we fit in this work (i.e., $L\gta10^4\lsun$) they measure $10\lta V_{\rm
  max}\lta30\kms$ which is, again, largely consistent with the range of
$V_{\rm max}$ we find. In this instance, the \citet{bovill11a} model produces
more bright, massive satellites than we see in the Local Group. They discuss
this in \citet{bovill11b} as the ``missing bright satellite''
problem. However, for the systems with comparable luminosities, there is
significant overlap in their masses.

From these comparisons, we are content that the best fit profiles to the MW,
M31 and total Local Group dSph we have derived are not hugely at odds with
predictions from simulations. Some tension remains at the higher end of the
subhalo mass range, as the simulations we compare with identify at least a few
subhalos with greater $V_{\rm max}$ than are compatible with observations.

\section{The mass-to-light ratios and circular velocities of Local
  Group dSphs}
\label{sect:outliers}

\begin{deluxetable}{lccc}
\tabletypesize{\footnotesize}
\tablecolumns{4} 
\tablewidth{0pt}
\tablecaption{The masses, mass-to-light ratios, and $V_{\rm max}$ values for
  Local Group dSphs as derived in this work}
\tablehead{
\colhead{Name} &   \colhead{$M_{\rm half}$} & \colhead{$[M/L]_{\rm half}$}
&$V_{c,1/2}$ \\ & \colhead{($\times10^7\msun$)}& \colhead{$(\msun/\lsun$)} & \colhead{$(\kms)$} \\ }
\startdata
AndI&  5.05$\pm1.35$ &  22.4$\pm8.5$ &  16.1$\pm4.4$ \\
AndII&  3.31$\pm0.7$ &  27.6$\pm10.8$ &  12.3$\pm2.6$ \\
AndIII&  1.95$\pm0.46$ &  39.0$\pm12.9$ &  14.7$\pm3.7$ \\
AndV&  2.30$\pm0.40$ &  78.0$\pm19.5$ &  16.6$\pm3.3$ \\
AndVI&  4.67$\pm0.9$ &  27.5$^{+7.61}_{-6.85}$ &  19.5$^{+4.2}_{-3.9}$ \\
AndVII&  7.61$\pm0.9$ &   9.0$\pm1.6$ &  20.5$\pm2.7$ \\
AndIX&  2.25$\pm0.7$ & 302$\pm132$ &  17.2$\pm6.0$ \\
AndX&  0.46$\pm0.2$ &  61.2$^{+52}_{-49}$ &  10.1$^{+5.1}_{-4.3}$ \\
AndXI&  0.41$^{+0.3}_{-0.2}$ & 165.5$^{+  196}_{-  142}$ &  12.0$^{+11.0}_{-8.1}$ \\
AndXIII&  0.26$^{+0.22}_{-0.16}$ & 126.6$^{+  153}_{-  108}$ &  9.2$^{+10.1}_{-6.4}$ \\
AndXIV&  0.42$\pm0.2$ &  41.6$\pm28.2$&   8.4$\pm5.2$ \\
AndXV&  0.22$\pm0.11$ &   9.0$^{+7.1}_{-7.0}$ &   6.3$^{+3.4}_{-3.3}$ \\
AndXVI&  0.24$^{+0.08}_{-0.06}$ &   11.6$^{+3.9}_{-2.9}$ &   8.8$^{+3.2}_{-2.7}$ \\
AndXVII&  0.13$^{+0.33}_{-0.19}$ &  12.82$^{+   44.79}_{-   26.38}$ &   4.5$^{+11.2}_{-4.5}$ \\
AndXVIII&  1.5$\pm0.5$ &  44.8$\pm27.1$ &  15.3$\pm6.1$ \\
AndXIX&  2.7$^{+1.9}_{-1.2}$ & 118.0$^{+  124.2}_{-   85.2}$ &   7.4$^{+6.6}_{-3.8}$ \\
AndXX&  0.30$^{+0.28}_{-0.17}$ & 213.0$^{+  282.2}_{-  171.0}$ &  11.2$^{+12.0}_{-7.0}$ \\
AndXXI&  1.2$^{+0.5}_{-0.4}$ &  29.8$^{+18.7}_{-   16.5}$ &   7.1$^{+3.1}_{-2.7}$ \\
AndXXII&  0.10$^{+0.11}_{-0.08}$ &  69.7$^{+  102.2}_{-   82.4}$ &   4.4$^{+4.7}_{-3.9}$ \\
AndXXIII&  3.4$^{+0.8}_{-0.7}$ &  68.4$^{+   46.4}_{-   46.1}$ &  11.2$^{+2.8}_{-2.6}$ \\
AndXXV&  0.33$^{+0.19}_{-0.18}$ &  10.2$^{+   10.0}_{-9.5}$ &   4.7$^{+2.9}_{-2.6}$ \\
AndXXVI&  0.83$^{+0.72}_{-0.46}$ & 279.8$^{+  383}_{-  277.6}$ &  13.7$^{+   15.7}_{-9.6}$ \\
AndXXVIII&  0.53$^{+0.36}_{-0.27}$ &  50.5$^{+   51.0}_{-   39.1}$ &  10.4$^{+7.7}_{-5.8}$ \\
AndXXIX&  0.68$^{+0.23}_{-0.22}$ &  67.8$^{+   38.6}_{-   37.5}$ &   9.0$^{+3.4}_{-3.2}$ \\
AndXXX&  2.1$^{+2.0}_{-1.2}$ & 300.0$^{+  433.1}_{-  302.3}$ &  18.6$^{+17.7}_{-11.4}$ \\
Scl&  1.3$\pm0.3$ &  18.2$\pm9.8$ &  14.5$\pm3.9$ \\
LeoT&  0.58$\pm0.22$ & 196.9$\pm120.0$ &  11.8$\pm5.1$ \\
UMaI&  2.6$^{+1.2}_{-1.1}$ &3731$^{+ 2577}_{- 2524}$ &  18.8$^{+8.9}_{-8.5}$ \\
LeoIV&  0.13$\pm0.10$ & 299.1$\pm54.2$ &   5.2$\pm4.0$ \\
Com&  0.09$\pm0.03$ & 510.8$pm309.0$ &   7.3$\pm2.2$ \\
CVnII&  0.09$\pm0.03$ & 229.9$^{+  158}_{-  152}$ &   7.3$^{+3.0}_{-2.6}$ \\
LeoI&  1.2$\pm0.3$ &   7.1$\pm3.3$ &  14.5$\pm3.5$ \\
LeoII&  0.38$\pm0.07$ &  12.9$\pm5.2$ &  10.4$\pm2.2$ \\
Car&  0.6$\pm0.2$ &  50.7$\pm28.9$ &  10.4$\pm3.0$ \\
UMi&  1.5$\pm0.3$ & 146.6$\pm76.5$ &  15.0$\pm2.9$ \\
Dra&  0.94$\pm0.18$ &  69.7$\pm22.0$ &  14.4$\pm3.0$ \\
For&  5.3$\pm0.6$ &   7.6$\pm2.5$ &  18.5$\pm2.4$ \\
Sex&  2.5$\pm0.71$ & 120.4$\pm74.4$ &  12.5$\pm4.2$ \\
Boo&  0.30$^{+0.08}_{-0.06}$ & 198.0$^{+   83.4}_{-   69.1}$ &   7.3$^{+2.0}_{-1.6}$ \\
CVnI&  1.9$\pm0.2$ & 164.3$\pm31.2$ &  12.0$\pm1.4$ \\
Herc&  0.26$^{+0.11}_{-0.10}$ & 145.6$^{+   95.8}_{-   89.7}$ &   5.8$^{+2.8}_{-2.4}$ \\
LeoV&  0.04$\pm0.05$ & 197.5$\pm339.1$ &   3.8$\pm4.4$ \\
Wil1&  0.03$\pm0.02$ & 536.2$\pm613.8$ &   6.8$\pm4.6$ \\
UmaII&  0.26$\pm0.10$ &1319.1$\pm961$ &   9.0$\pm3.9$ \\
Seg&  0.02$\pm0.01$ &1374.7$^{+ 1453.47}_{- 1234.16}$ &   5.8$^{+3.9}_{-2.8}$ \\
Seg2&  0.02$\pm0.02$ & 536.4$\pm588.7$ &   4.1$^{+0.9}_{-4.1}$ \\
\enddata
\label{tab:ml}
\end{deluxetable}

\begin{figure}
  \begin{center}
    \includegraphics[angle=0,width=0.9\hsize]{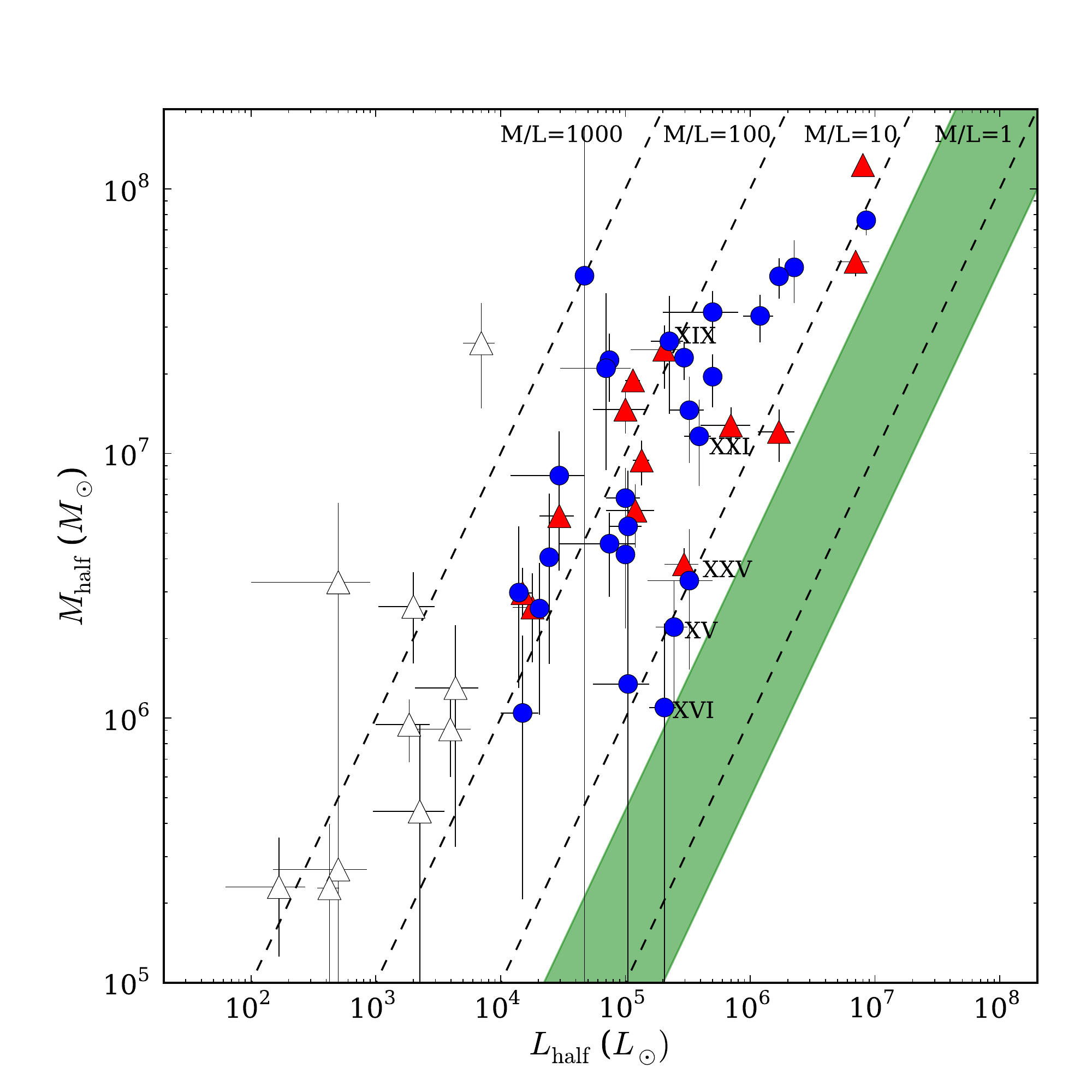}
    \caption{Mass within the half-light radius ($M_{\rm half}$) as a function
      of luminosity within the half-light radius ($L_{\rm half}$) for Local
      Group dSphs. Points are colour coded as in previous figures. The dashed
      lines represent mass-to-light ratios of $[M/L]_{\rm half}=1,10,100$ and
      $1000\msun/\lsun$. The green shaded region indicates the parameter space
      typically inhabited by simple stellar systems (i.e., those without dark
      matter). It is interesting to note that there are a number of M31
      objects, namely And XV and XXV, that are consistent with this
      regime. }
 \label{fig:ML}
 \end{center}
 \end{figure}

In Fig.~\ref{fig:ML}, we plot the masses contained within the half-light radii
($M_{\rm half}$) of all the Local Group dSphs as a function of their
luminosity within the half-light radius ($L_{\rm half}$), where the points are
colour coded as in previous figures. The values themselves can be found in
Table~\ref{tab:ml}. Additionally we overplot lines of constant mass-to-light ratio (with
$[M/L]_{\rm half}=1,10,100$ and $1000$). It can be seen that the majority of
these objects (including two of our outliers, And XIX and XXI, labeled in
plot) have $[M/L]_{\rm half}\gta10$, indicating that their dynamical masses
are much higher than can feasibly be explained by the mass of their baryons
alone (although see recent work in predicting the velocity dispersions and
mass-to-light ratios of M31 dSphs using MOND, without dark matter by
\citealt{mcgaugh13}). This is typically ascribed to the presence of dark
matter halos in these objects, whose mass dominates that of their baryons. The
green shaded region in this plot represents the parameter space in this
framework typically inhabited by globular cluster systems of the MW
\citep{rejkuba07}, whose masses can be explained by their stellar content
alone, without invoking dark matter.

Interestingly, we see a few objects on this plot whose mass-to-light ratios
are consistent (within $1\sigma$ uncertainties) with those of the Galactic
globular clusters, suggesting that they possess little or no dark matter. In a
couple of cases, the very large uncertainties on current measurements mean
that this overlap is not significant, and will likely disappear with future
observations. But there are two objects, And XV and XXV, that are particular
noteworthy. The masses of And XV and XXV are derived from sample sizes of
$\sim30$ stars. The potential implication of this is that these galaxies
contain very little dark matter, which would be quite unexpected for objects
of their sizes and luminosities.

\begin{figure*}
  \begin{center}
    \includegraphics[angle=0,width=0.9\hsize]{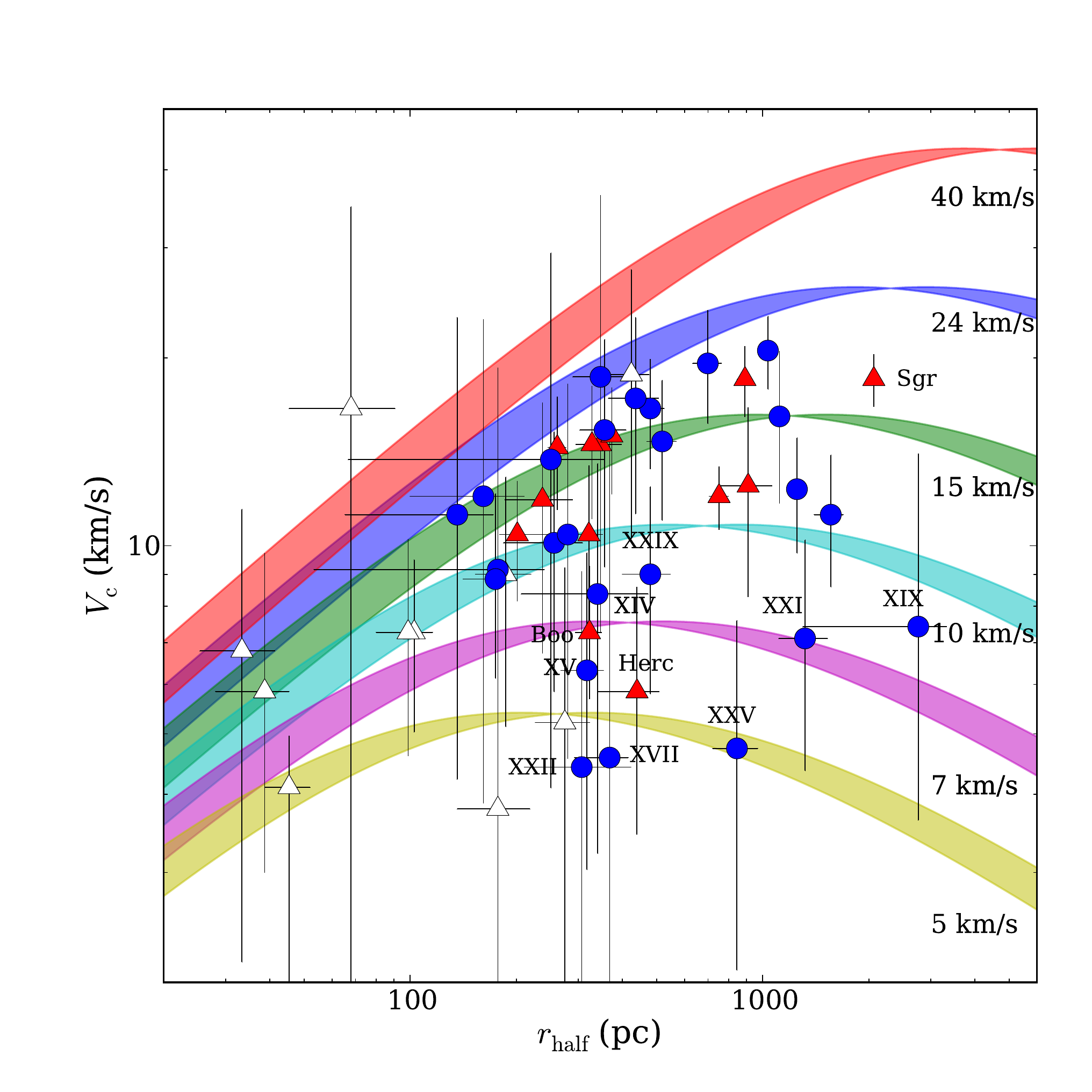}
    \caption{ Circular velocities within the half-light radius($V_{c,1/2}$) for
      Local Group dSphs, as derived from their velocity dispersions. The
      shaded lines are circular velocity profiles of subhalos from the
      Aquarius simulations (taken from \citealt{kolchin12}), and are labeled
      with their maximum circular velocities. Subhalos with maximum circular
      velocities below $10\kms$ (i.e. profiles below the cyan shaded curve)
      are not thought to be massive enough to efficiently cool their hydrogen
      and form stars. A number of dSphs (particularly Herc and And XXV) have
      circular velocities that see them preferentially residing in halos that fall
      below this low mass cut-off for star formation.}
 \label{fig:vmax}
 \end{center}
 \end{figure*}

 In Fig.~\ref{fig:vmax} we plot the circular velocities measured within
  the half-light radius (a good proxy for the central mass of these objects)
  of the Local Group dSphs as a function of their half-light radii. We derive
  $V_{c, 1/2}$ from the measured masses within the half-light radius using the
  relationship between circular velocity and mass:
  $V_{c,1/2}=\sqrt{\frac{GM_{\rm half}}{r_{\rm half}}}$. These values are
  tabulated in Table~\ref{tab:ml}. The shaded regions overplotted represent
  the circular velocity profiles of Aquarius subhalos (taken from
  \citealt{kolchin12}) and are labeled with their maximum circular velocity in
  each case. Subhalos with maximum circular velocities below $10\kms$
  (i.e. below the cyan profile in Fig.~\ref{fig:vmax}) are proposed to be too
  low mass to form luminous galaxies, as their star formation is highly
  suppressed due to inefficient gas cooling, causing them to remain
  essentially dark ($V_{\rm max, limit}\sim10\kms$, \citealt{koposov09}).  The
  red and blue curves in the figure are representative of the $\sim10$
  most-massive subhalos seen in DM only simulations where we would naively
  expect the most luminous dwarf galaxies in the Local Group to reside. When
  including the M31 dSphs, we see that there are now a number of systems that
  may be consistent with living in such massive halos. However, many of these
  systems are the less luminous objects ($-6>M_V>-8$), where our measurement
  uncertainties are large. For the brighter M31 dSphs whose velocity
  dispersions (and hence, masses) are well resolved, there are only 2 objects
  (And VI and And VII) that may inhabit halos with maximum circular velocities
  of $24\kms$ or greater. As such, the TBTF problem would seem to be present
  in Andromeda as well as the Milky Way.

The previously discussed outliers from C13 and T12 (Hercules, And XIV, XV, XVI, XIX, XXI
and XXV) are all labeled in Fig.~\ref{fig:vmax}, as is the MW dSph,
Bo\"otes I (Boo I). These objects again stand out as
they fall tentatively shy of the pre-reionization star formation
threshold. If their halos have always been so low mass, they should never have
been able to form stars. Given the large uncertainties, all but 4 of these
outliers are (just) consistent with this lower limit,and thus, not of great
concern. But And XXII and XXV in M31, and Herc and Boo I in the MW, all fall
below this threshold, even when taking their uncertainties into account. This
implies that, in order for us to observe these systems now, their masses must
have been higher in the past, and have been reduced by some physical process
during their evolution. We discuss this further in \S~\ref{sect:discussion}.

Combined, these low mass-to-light ratios and lower than predicted masses,
highlight the ongoing tensions between observations and theory. It is clear
that, if the predictions from the $\Lambda$CDM paradigm are to be reconciled with our
observations, we must investigate avenues that can lower the masses of dark
matter subhalos over the course of their cosmic evolution. In the next
section, we discuss numerous possibilities for this that have been put forth
recently, and comment on their ability to reproduce our findings within the
Local Group.

\section{Explaining the masses of the Local Group dSphs - observations
  vs. theory}
\label{sect:discussion}

From the analysis in \S~\ref{sect:results} and \S~\ref{sect:outliers} we still
see some discrepancies between the masses of subhalos in simulations, and the
masses we infer from observations of the subhalos within the Local Group. At
face value, it seems that we expect to observe luminous satellites around MW
mass halos with higher central masses than we do. One explanation for this
missing massive satellite problem could simply be that at these low halo
masses ($M_{\rm halo}<10^{10}\msun$) star formation becomes increasingly
stochastic, so that the luminosity of a subhalo does not necessarily correlate
with the mass of the subhalo \citep{kuhlen12,kuhlen13}. Other solutions appeal
to physical processes affecting the evolution of dwarf galaxies, and can be
broadly assigned to three categories: The effect of tidal interactions with
the host galaxy, the effect of stellar feedback on the mass profiles of
galaxies and the true mass of the host system. Finally, some have also
appealed to the modification of the current cosmological paradigm,
$\Lambda$CDM, either via the properties of the dark matter itself (e.g. warm
dark matter, \citealt{anderhalden13} or self interacting dark matter,
\citealt{rocha13,zavala13}), or via the modification of Newtonian gravity
(i.e. MOND, \citealt{mcgaugh13}) to remove the need for dark matter
altogether. In the subsequent sections, we discuss the physical processes that
might be responsible for lowering the central masses of the whole Local Group
population, and why this effect might be more pronounced in some objects
(i.e. Boo I, Herc, And XIX, XXI and And XXV) than others.

\subsection{Tides}

The notion that the lower central masses observed in the outlying dSphs is the
result of some physical process that has moved them away from a more typical
mass is appealing, and the possibility has been briefly discussed in other
works, particularly \citet{collins11b} and C13. There, the authors point out
that dSphs that are more extended at a given luminosity, such as the extreme
outliers, And XIX, XXI and XXV, also tend to have lower central masses. One of
the MW low mass outliers, the MW Herc object, is also already thought to be a
tidally disrupting system, transitioning into a stellar stream
\citep{martin10}. 

Outside of the Local Group, a number of tidally disrupting dwarf galaxy
systems have recently been observed whose half-light radii are also much more
extended than would be expected from the $r_{\rm half}-L$ relation
\citep{brasseur11b}, resulting in very low surface brightnesses. These include
NCG 4449B ($M_V=-13.4$ and $r_{\rm half}=2.7$~kpc,
\citealt{rich12,delgado12}), and the Hydra dwarf galaxy, HCC-087 ($M_V=-11.6$
and $r_{\rm half}=3.1$~kpc, \citealt{koch12}). Their closest analog within the
Local Group is And XIX ($M_V=-9.3$, $r_{\rm half}=2.1^{+1.0}_{-0.4}\kpc$,
Martin et al. in prep.) making it an outlier to the \citet{brasseur11b} relation,
as well as falling below the mass expectation for a galaxy of its radial
scale.  Taking these examples into account, perhaps
these properties (low surface brightness and/or low mass) are indicators of a
system undergoing significant tidal interaction with its host.

 Mass loss of subhalos from interactions with their hosts has also been
  studied in numerical models, both in a dark matter only context (e.g.,
  \citealt{tormen98,klypin99,ghigna00,hayashi03,zentner03,kravtsov04,kazantzidis04a})
  and with the inclusion of baryons (e.g.,
  \citealt{penarrubia10,donghia10,zolotov12,brooks12}). In all cases, as these
  systems orbit their host, their dark matter is stripped, lowering their
  densities and masses at all radii. After the dark matter is removed, the
  stars reach a dynamic equilibrium with their lower density potential,
  causing a drop in the central mass. In simulations where baryonic physics
  are included, the mass losses from subhalos as a result of tidal
  interactions with a host are more pronounced than for the dark matter
  only case. Further, the size of the mass reduction increases with earlier infall
  times and more radial orbits. In \citet{zolotov12}, they demonstrated that a
  subhalo accreted at $z>6$~Gyr in an SPH simulation would experience a
  greater reduction in its mass than is seen with a dark matter only set
  up. Similarly, the mass of subhalos on radial orbits in the SPH simulation
  also experience a more significant drop in mass than their dark matter only
  counterparts. In all cases, the presence of a massive baryonic disk in the
  host galaxy (such as those hosted by the Galaxy and M31) reduces the masses
  of the satellite population at a much greater rate than in the dark matter
  only case.

One could therefore argue that the outliers seen in this study, such as
Hercules, And XIX, XXI and XXV, may have fallen in to their host galaxies
earlier, and onto more radial orbits where they interact more significantly
with their host, leading to a more pronounced mass loss. It is difficult to
properly model the orbital properties of these objects, but recent work
by \citet{watkins13} modelled the orbital properties of M31 dSphs by combining
the timing argument with phase-space distribution functions. This work found
no evidence to suggest that the M31 outliers are on very radial orbits, nor do
they seem to have experienced particularly close passages with M31 itself,
perhaps ruling out this option.

A prime example of a tidally disrupting dSph within the MW is the Sgr
dSph. This object is currently undergoing violent tidal disruption, yet it has
a velocity dispersion that is entirely consistent with the best fit NFW and
cored mass profiles to both the MW alone and to the full Local Group, perhaps
arguing against the mechanism we have outlined above. However, Sgr is
currently near the pericenter of its orbit, only $\sim20$ kpc from the
Galactic center \citep{law10}. The outliers we refer to are located further
out ($D_{host}>70\kpc$ for all outliers, \citealt{martin10,koposov11,conn13}),
and so we do not expect them to be currently experiencing significant tidal
distortions, rather that their past interactions with their host have removed
more mass from their centers than their more `typical' counterparts.

In summary, numerical models have demonstrated that tidal mechanisms are able
to lower the masses of dSphs, and could explain the lower than expected masses
of the Local Group outliers, Herc, And XIV, XV, XVI, XIX, XXI and XXV if they
have experienced more significant past interactions with their host.

\subsection{Feedback from star formation and supernova}

For many years, kinematic studies of low surface brightness galaxies have
shown that the mass profiles of these objects are less centrally dense than
expected. They are more compatible with flatter, cored halo functions, rather
than the cuspier NFW profiles seen in simulations
(e.g. \citealt{flores94,deblok02,deblok03,deblok05}). Many have argued that
this is a result of bursty, energetic star formation and supernova
(SN) within these galaxies. These processes drive mass out from the center of the halo, flattening the high
density cusp into a lower density core, leading to a lower central mass than
predicted by pure dark matter simulations
(e.g. \citealt{navarro96b,dekel03,read05,mashchenko06,pontzen12,governato12,maccio12}). Could
the lower than expected central masses of the Local Group dSphs also be caused
by feedback?

\citet{zolotov12} and \citet{brooks12} compared a dark matter only simulation
with a smooth particle hydrodynamic (SPH) simulation of a MW type galaxy in a
cosmological context to see whether the inclusion of baryons and feedback in
the latter can produce satellite galaxies with lower central masses and
densities. For galaxies with a stellar mass $M_*>10^7\msun$ ($M_V\lta-12$) at
the time of infall, feedback can reduce the central mass of dSph
galaxies. Below this mass, the galaxies have an insufficient total mass to
retain enough gas beyond reionization to continue with the significant, bursty
star formation processes required to remove mass from their centers. The Local
Group outliers discussed above have $M_V\gta-10$, so unless they have been
significantly tidally stripped by their hosts after falling in, i.e.,
experienced {\it total} (dark matter plus baryonic) mass losses of greater
than $\sim90\%$ \citep{penarrubia08b}, feedback cannot explain their current
masses. This is supported by the findings of \citet{garrison13}, where they
model the dynamical effect of SN feedback on the mass distribution of dark
matter halos. To match the current observed central masses in MW dSph
galaxies, one would need to deposit 100\% of the energy resulting from $40,000$
SN directly to the dark matter halos, which is greater than the expected total
number of SN to have ever occurred in the majority of these systems. The work
of \citet{penarrubia12} also support the findings of these works, namely that
fainter dSphs should not be able to significantly lower their central masses
via feedback, and if they were able, they would serve to exacerbate the
missing satellite problem.

In \citet{dicintio13}, they also study the effect of feedback in subhalos on
their mass profiles, using a suite of galaxies from the MaGICC project
\citep{stinson10,stinson13}. Their findings show that the mass of stars formed
per halo mass is the most important factor for the shaping of the central mass
profiles of galaxies. Objects with stellar-to-halo mass ratios of $M_*/M_{\rm
  halo}\lta0.01$ are not able to alter their dark matter distributions, but as
this fraction increases, so does the ability to flatten their central mass
profiles. They find that this process is maximally efficient in galaxies with
$M_*\approx10^{8.5}\msun$, and below this, the central dark matter slope
increases once more. As such, their findings are similar to those of
\citet{zolotov12}. This seeming consensus on the amount of baryons
  required to efficiently reshape the dark matter mass profile of a dwarf
  galaxies via feedback means that, in principle, based on their current luminosities only 4
  of the dSphs discussed in this work (And II, And VII, Fornax,
  and Leo I, \citealt{mcconnachie12}) have enough explosive energy at
  their disposal to reduce their central densities with feedback alone. For the
  remaining MW and M31 objects, another mechanism, such as tides, would need
  to be invoked to explain their low masses.

\subsection{Host mass}

 From a plethora of works
  (e.g. \citealt{moore99,ghigna00,kravtsov04b,zentner05,vandenbosch05,zheng05,giocolo08,springel08}),
  we know that the number of subhalos within a host halo, scales with the mass
  of the host halo itself. Therefore, when comparing the mass of halos we
observe within the MW or Andromeda with those found in simulations, it is
important that we select a simulated galaxy of the same mass. Unfortunately,
in the case of both M31 and the MW, the total masses of these systems are
actually quite uncertain, ranging from $\sim0.7-2.7\times10^{12}\msun$ for the
Milky Way (e.g., \citealt{wilkinson99,xue08,li08,watkins10,piffl13}) and
$\sim0.8-2.2\times10^{12}\msun$ (e.g.,
\citealt{evans00a,evans00b,li08,guo10,watkins10}), making this difficult. From
the point of view of abundance matching, a galaxy as luminous as the MW should
be hosted by a halo with an even higher mass than these estimates
($\sim3\times10^{12}\msun$, \citealt{behroozi13}), which could imply that our
Galaxy is a significant outlier when compared with the bulk of the galaxies
within the Universe.

\citet{veraciro13} investigated the effect of varying the mass of the host
halo on both the number and dynamics of subhalos using the Aquarius
simulations \citep{springel08}. They found that they were able to match both
these quantities when using a simulated halo whose mass was consistent with
the lower bound of observational constraints for the MW,
$8\times10^{11}\msun$. This immediately eliminates the TBTF problem, as the
most massive simulated subhalos have $V_{\rm
  max}\lta25\kms$. \citet{dicintio12} also find that they can match the number
and dynamics of MW satellites using halos from the CLUES simulations with
masses of $5-7\times10^{11}\msun$ \citep{gottlober10,dicintio12}. Thus,
  if the virial mass of the MW is at the lower end of current observational
  estimates, the masses we measure for its subhalos would be much more inline
  with predictions from numerical simulations. However, it is worth noting
  that such a low mass for the MW would lessen the probability of hosting
  massive satellites like the LMC and SMC, and may just replace the missing
  massive satellite problem with a found massive satellite problem
  (e.g. \citealt{kolchin11b,busha11}). It is also in conflict with a recent
  estimate of the mass of the MW from \citet{kolchin13} who use the Aquarius
  simulations to demonstrate that the 3D space motion of the Leo I dSph puts a
  lower limit of $\sim1\times10^{12}\msun$ on the mass of the MW at a
  confidence level of 95\%. \citet{veraciro13} also find that the Andromeda
satellites can be best matched if the host mass is $1.77\times10^{12}\msun$,
implying that the mass of M31 is roughly twice that of the MW. This value is
compatible with the best estimate for the Andromeda mass when using the full
dwarf galaxy satellite population as a tracer (L. Watkins, private
communication). If Andromeda is more massive than the MW, it could explain
  the fact that M31 has more (in number terms) massive non-dSph satellites
  (NGC147, NGC 185, NGC 205, M32 and M33) than the MW (LMC and SMC). It may
  also explain the low mass outliers, particularly the statistically
  significant And XIX, XXI and XXV. In particular, the very low value of
  $V_{c,1/2}$ we derive for And XXV strongly indicates that it is currently
  residing in a dark matter subhalo with a maximum circular velocity below the
  mass limit expected for luminous galaxy formation, suggesting it has
  experienced some physical process that has lowered its mass significantly
  over the course of its evolution. A more massive host would imply that the
  dSphs (which are more susceptible to tidal disruption than the more massive
  satellites) within this system have experienced greater tidal forces over
  the course of their evolution, which could lower their masses below the SF
  threshold of $10-15\kms$ \citep{penarrubia08b,koposov09}.

Precisely pinning down the correct viral masses of the MW and M31 is clearly
an important step towards better understanding the masses of the dwarf
galaxies we observe within the Local Group in a cosmological context. Without
precise mass estimates, it is difficult for us to quantify discrepancies with
theoretical expectations, like those of the TBTF problem, and might help us to
explain the differences between the masses of dSphs we see around M31 and the MW.

\section{Conclusions}
\label{sect:conclusions}

The relatively high dark-to-stellar mass ratios of dSph galaxies single them
out as excellent probes of the behaviour of dark matter on the smallest of
scales. Comparisons of the masses of the MW dSphs with expectations from
cosmological simulations have revealed several discrepancies, most notably the
issue of cuspy vs. cored central densities and a dearth of luminous high mass
subhalos around the MW compared with dark matter only simulations (the `too
big to fail' problem). In this paper we have expanded these analyses by
including the dSph satellites of M31 in the comparisons.

We revisit the notion that all dSph are embedded in dark matter halos that
follow a universal density profile in their centers \citep{walker09b} by
fitting NFW and cored mass profiles to the full sample of MW and M31 dSphs in
$r_{\rm half}-\sigma_v$ space. We find that no singular profile provides a
good fit to the data, but that their masses are instead described by a range
of halo profiles with a well defined scatter as a function of half light
radius. We find that when comparing fits for solely MW dSphs to solely M31
dSphs, the latter prefer significantly lower masses for a given size than the
former. We demonstrate that this offset is driven by 3 low mass outliers in
M31, whose half-light radius place them in a region of parameter space with
very few MW dSphs for comparison (And XIX, XXI, and XXV). Once these outliers
are removed, we find that the two populations agree exceptionally well,
following mass profiles with similar average values of $V_{\rm max}$ and
$R_S$, and a scatter in mass that equates to $\sim50\%$ of the average mass at
any specific radius.

We also derive the $V_{c, 1/2}$ values for each dSph directly from their
velocity dispersions and find them to be in good agreement with our fitted
ranges, with the exception of the 3 excluded outliers. Further inspection of
these values, plus the mass-to-light ratios of the population reveal a number
of interestingly low mass systems. The most significant of these are And XV,
XIX, XXI and XXV from M31 and Herc and Boo I from the MW. In particular, the
central mass of And XXV is so low that if it had always been this way, it
would never have formed stars. And yet, now; it is clearly luminous. By
comparing the properties of these objects with those of observed tidally
disrupting dwarf galaxies, we postulate that tides are a candidate mechanism
for lowering the masses of these objects, especially when combined with
stellar feedback at early epochs that can reduce the central masses of these
galaxies before they fall into their host.

When comparing the computed values of $V_{c,1/2}$ from our observed sample
with the circular velocity profiles of subhalos within simulations, we still
see an offset between the most massive simulated subhalos, and the most
massive dSphs in both the MW and M31, described as the TBTF problem. We argue
that, as this problem was defined via comparisons with a dark matter only
simulation of a MW type halo that neglects the effects on the mass profiles of
dSphs from baryonic processes (such as feedback and tides), and may not be
directly comparable to the MW and M31 (given the uncertainties on
observational measurements of their masses), it is difficult to quantify how
serious or significant this problem truly is. By running simulations with
baryonic processes included, and precisely determining the masses of the MW
and M31, we will better be able to assess whether there is truly a missing
massive satellite problem.  As such, the masses of Local Group dSphs should be
thought of as a constraint for more complex simulations that include a wide
range of physical processes that are not currently well understood.

\section*{Acknowledgments}

We would like to thank the anonymous referee for their helpful comments and
suggestions, as well as a thorough discussion surrounding the `Too Big To
Fail' problem.

The research leading to these results has received funding from the European
Research Council under the European Union's Seventh Framework Programme (FP 7)
ERC Grant Agreement n. [321035].

R.I. gratefully
acknowledges support from the Agence Nationale de la Recherche though
the grant POMMME (ANR 09-BLAN-0228).

G.F.L. and N.B. gratefully acknowledge financial support for his
ARC Future Fellowship (FT100100268) and through the
award of an ARC Discovery Project (DP110100678).

A.K. thanks the Deutsche Forschungsgemeinschaft for funding from Emmy-Noether
grant Ko 4161/1.

\bibliography{mnemonic,michelle}{} \bibliographystyle{apj}

\end{document}